\documentclass[aps, prb,
twocolumn, notitlepage,
superscriptaddress,floatfix,nofootinbib]{revtex4-2}

\usepackage{amsmath}
\usepackage{amsfonts}
\usepackage{amsthm}
\usepackage{amssymb}

\usepackage{bm}
\usepackage{physics}

\usepackage{graphicx}
\usepackage{xcolor}
\usepackage[caption=false]{subfig}

\usepackage[breaklinks=true,colorlinks,citecolor=blue,linkcolor=red,urlcolor=blue]{hyperref}

\newcommand{\ii}{\mathrm{i}}
\newcommand{\ee}{\mathrm{e}}
\newcommand{\ce}{\varepsilon}

\newcommand{\fig}{Fig.}

\newcommand{\eq}{Eq.}

\begin{document}
\title{Dissipative realization of Kondo models}
\author{Martino Stefanini}
\affiliation{Institut f\"ur Physik, Johannes Gutenberg-Universit\"at Mainz, D-55099 Mainz, Germany}
\author{Yi-Fan Qu}
\email{Corresponding author: yifaqu@phys.ethz.ch}
\affiliation{Institute for Theoretical Physics, ETH Z\"urich, 8093 Zurich, Switzerland}
\author{Tilman Esslinger}
\affiliation{Institute for Quantum Electronics \& Quantum Center, ETH Zurich, 8093 Zurich, Switzerland}
\author{Sarang Gopalakrishnan}
\affiliation{Department of Electrical and Computer Engineering, Princeton University, Princeton, New Jersey 08544, USA}
\author{Eugene Demler}
\affiliation{Institute for Theoretical Physics, ETH Z\"urich, 8093 Zurich, Switzerland}
\author{Jamir Marino}
\affiliation{Institut f\"ur Physik, Johannes Gutenberg-Universit\"at Mainz, D-55099 Mainz, Germany}

\date{\today}

\begin{abstract}
    We demonstrate that the Kondo effect can be induced through non-linear dissipative channels, without requiring any coherent interaction on the impurity site. Specifically, we consider a reservoir of noninteracting fermions that can hop on a few impurity sites that are subjected to strong two-body losses. In the simplest case of a single lossy site, we recover the Anderson impurity model in the regime of infinite repulsion, with a small residual dissipation as a perturbation.
    While the Anderson model gives rise to the Kondo effect, this residual dissipation competes with it, offering an instance of a nonlinear dissipative impurity where the interplay between coherent and incoherent dynamics emerges from the same underlying physical process.
    We further outline how this dissipative engineering scheme can be extended to two or more lossy sites, realizing generalizations of the Kondo model with spin 1 or higher.
    Our results suggest alternative implementations of Kondo models using ultracold atoms in transport experiments, where localized dissipation can be naturally introduced, and the Kondo effect observed through conductance measurements.
\end{abstract}
\maketitle

\section{Introduction}
The Kondo effect \cite{Hewson} is one of the simplest and most iconic phenomena in the physics of strongly correlated systems. It emerges when an interacting impurity exchanges particles with a gapless fermionic reservoir. The hybridization of the impurity levels with the bath's states causes the emergence of a very narrow many-body resonance (the Kondo, or Abrikosov-Suhl resonance) pinned at the chemical potential of the reservoir, whose properties dominate the low-energy physics and lead to a number of fascinating phenomena, such as universal scaling behavior of thermodynamics quantities \cite{Hewson, KondoReview,Coleman,Kondo,ResistanceMinimum,ClogstonResistanceMinimum,Abrikisov,Suhl,BlandinNozieres,AndersonImpurityModel,SchriefferWolff,CoqblinShrieffer} in impure metals, and an almost perfect conductance through the impurity in quantum dots \cite{ZeroBiasAnomalyPRL,GlazmanRaikh,transport_aim,PerfectGKondo,Review_quantum_dots,Nature_kondo_quantum_dot,Science_tunable_Kondo}.
From a theoretical perspective, the Kondo effect is a source of enduring interest as it showcases how strongly correlated behavior can emerge from simple ingredients.
\par With the advent of quantum simulation with ultracold atomic gases, new possibilities have arisen for the study of strongly correlated physics in regimes that would be otherwise inaccessible to traditional solid state setups, and consequently there have been many proposals for realizing the Kondo effect \cite{expKondoFalco04,expKondoDuan04,expKondoRecati05,expKondoParedes05,expKondoOrth08,expKondoSarang10,expKondoGorshkov10,expKondoBauer13,expKondoNishida13,expKondoNishida16,expKondoIsaev15,expKondoKuzmenko15,expKondoNakagawa15,expKondoZhang16,PRB_Anisotropic_Kondo}. Such realization would be desirable for accessing less understood properties of the Kondo model and its relatives, such as their nonequilibrium dynamics, including the spreading of correlations in real space---i.e., the formation of the Kondo screening cloud, which has eluded measurement in conventional platforms until recently \cite{KondoCloud}. Moreover, realizing the Kondo models would provide a stepping stone for implementing more complicated models that are harder to realize in solid state setups, such as high-spin and multi-channel Kondo models \cite{Bickers,Hewson,GogolinNersesyanTsvelik,Giamarchi,Cox_Zawadowski_exotic_Kondo,CoqblinShrieffer,NozieresBlandin,Hirst_ionic_model,Nature_2ck,PRB_Kiselev_two_level_Kondo, PRB_Kiselev_multistage_Kondo,PRL_Ludwig_Affleck_multichannel_Kondo,Thermodyn_multichannel_kondo}, or whose theoretical understanding is more limited, such as the Kondo lattice that is used to model heavy fermion compounds \cite{Hewson,Coleman,PRL_heavy_fermions,RMP_heavy_fermions}. Despite the numerous proposals, the Kondo effect is yet to be observed with ultracold atoms. Only the first steps in achieving a Kondo lattice configuration have been performed so far \cite{PRL_Experimental_FM_Kondo}.
\par In this work,
we show how a strong, localized two-body loss \cite{DipolarMolecules2BLoss, PRL_2BLoss_Zeno, DissipativeFermiHubbard,Bose_Hubb_2b_loss,Science_2b_loss} within a noninteracting fermionic gas can provide the correlations necessary to induce the Kondo effect. 
The significance of our results is twofold. On the one hand, we provide a physically motivated example of a \emph{nonlinear} dissipative impurity system, in which the many body nature of dissipation imprints correlations on the system. Indeed, we observe a competition between the incoherent effect of losses and the coherent, Kondo dynamics that they induce. The present work complements the growing corpus of literature that has been devoted to \emph{linear} impurities, i.e., featuring single-body losses \cite{FromlPRL,FromlPRB1,FromlPRB2,PRB_Alba_Carollo,PRA_Uchino,Esslinger_PRA_theory,Esslinger_PRL,Esslinger_PRL_23,Ott_PRL_bistability,Ott_localized_diss_BEC,lin_imp_Visuri,lin_imp_Alba,lin_imp_Wolff} or gains \cite{LocalizedFermionSource}, as well as local dephasing \cite{Ott_dissipative_imp,Dolgirev}. On the other hand, the present work points to a dissipative route to the implementation of both standard and more exotic Kondo models in ultracold atoms. An appealing feature of this route is that losses can be easily incorporated in the recently developed transport experiments with ultracold atoms \cite{Esslinger_JPhys,Esslinger_PRA_theory,Esslinger_PRL,Esslinger_PRL_23,Esslinger_PRX,Ott_PRL_negative_cond,Ott_PRL_bistability,Ott_dissipative_imp,Ott_localized_diss_BEC} that mimic the configuration of mesoscopic systems like quantum dots. This setup introduces the possibility of revealing the emergence of the Kondo effect in the same, well-established way of quantum dots, namely via conductance measurements. This approach is rather direct and may prove to be more sensitive than direct measurements of the spectral function through radio-frequency spectroscopy \cite{rfSpectroscopy}. 
\par We remark that our results regard the emergence of a typically Hamiltonian effect by means of dissipation. In this, we are distinguished from other recent works that use dissipation to introduce entirely new features, such as engineering of non-Hermitian versions of the Kondo model \cite{NHKondo_Ueda,hasegawaNHKondo,NHKondo_Andrei} and measurement-induced crossovers in continuously monitored quantum dots \cite{hasegawaNHKondo,KondoZenoSchiro}.
\par Our main finding is the characterization of the Kondo effect induced by a strong two-body loss localized on a single site connected to two reservoirs of noninteracting fermions. We argue that for an infinitely strong dissipation this system realizes the well-known Anderson impurity model (AIM) with infinite repulsion, and we derive the leading corrections to the Lindblad master equation for finite dissipation rate, which take the form of a residual two-body loss. We analyze the typical signatures of the Kondo effect (Kondo resonance in the spectral function, enhanced differential conductance at zero bias and suppressed decay of magnetization) both in dynamics and in the local steady state, and we observe a competition between the Kondo effect and the residual losses, with the latter suppressing the former as the dissipation rate is decreased. 
\par We conclude by describing how to realize higher-spin Kondo models by distributing the dissipation on more than one site, provided they are strongly coupled among themselves, and we briefly discuss the extension of the model to a multi-channel scenario. 
   
\section{Results and discussion}
\subsection{Model}
\begin{figure*}[ht]
    \centering
    \includegraphics[width=\linewidth]{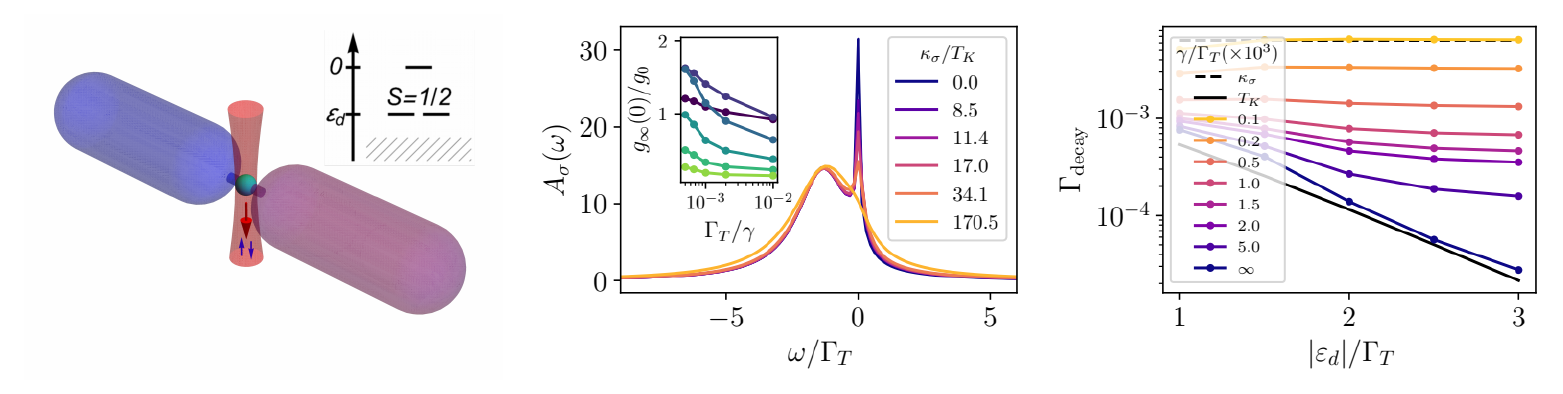}
    \caption{Using strong, localized two-body losses to realize the Anderson impurity model with infinite repulsion. Left panel: sketch of the simplest dissipative setup considered in the main text. The two tubes represent reservoirs of noninteracting, spinful fermions (possibly at different chemical potentials) that are allowed to tunnel to the central site, which is subjected to strong two-body losses. Inset: sketch of the dark states of the isolated impurity subjected to the two-body loss. The hatched area represents the dissipated states. Central panel: impurity spectral function in the local steady state, showing the smearing of the Kondo resonance as the effective dissipation is increased. The parameters are $\ce_d=-2\Gamma_T$ and $\Gamma_T=10^{-2}W$ (we take the half-bandwidth $W$ as our energy unit), for unbiased reservoirs at $\mu=0$ and zero initial temperature. Inset: zero-bias conductance across the impurity, normalized to $g_0=1/h$ ($h$ being Planck's constant). Lighter colors correspond to larger $\abs{\ce_d}/\Gamma_T\in\{0.5,\,1,\,1.5,\,2,\,2.5,\,3\}$. Right panel: decay rate of the impurity magnetization, showing a crossover from the Kondo regime $\Gamma_\text{decay}\sim T_K$ to the incoherent regime $\Gamma_\text{decay}\sim \kappa_\sigma$. In these plots $\Gamma_T=10^{-2}W$. In both the central and right panels we use the definition of $T_K$ reported in the main text, $T_K\equiv (\Gamma_T W/2)^{1/2}\exp[-\pi(\mu-\ce_d)/\Gamma_T]$.}
    \label{fig: aim}
\end{figure*}
We consider the system depicted in the leftmost part of \fig~\ref{fig: aim}, composed of two reservoirs (or leads) of noninteracting, spinful fermions that are connected to a central region (the “dot”) whose sites are subjected to a two body loss---whenever two opposite-spin fermions occupy one of the dissipative sites, they can both be lost from the system with a certain rate $\gamma$ \footnote{In principle one could include also one- and three-body losses, but these processes are weak in a noninteracting Fermi gas, and can be neglected with respect to the large two-body losses assumed here.}. We are going to comment on the many-site and many-leads scenario at the end of the paper. In the rest of this work we will focus on the simplest case of a single dot site hosting a single orbital, and coupled to two reservoirs. In the conditions common to experiments with ultracold atoms \cite{DipolarMolecules2BLoss, PRL_2BLoss_Zeno, DissipativeFermiHubbard,Esslinger_PRA_theory} losses are Markovian, and the dynamics of the system's density matrix $\rho(t)$ can be described by a Lindblad master equation ($\hbar=1$)
\begin{equation}
    \dv{}{t}\rho(t)=-\ii\comm{H}{\rho(t)}+\gamma\Big(L\rho(t)L^\dag-\frac{1}{2}\{L^\dag L,\rho(t)\}\Big)
\end{equation} 
with jump operator $L=d_\downarrow d_\uparrow$ (where $d_\sigma$ annihilates a fermion with spin $\sigma$ on the dot, with $\sigma\in\{\uparrow,\,\downarrow\}=\{+,\,-\}$). The Hamiltonian has the familiar form of a resonant level model \cite{HaugJauho}
\begin{equation}\label{eq: Hamiltonian}
    \begin{aligned}
        H=&H_d+H_\textup{leads}+H_\textup{tun} \\
        =&\ce_d\sum_\sigma d_{\sigma}^\dag d_{\sigma}+\sum_{p\sigma\alpha} \ce_{p\alpha} c_{p\sigma\alpha}^\dag c_{p\sigma\alpha} \\
    &+\sum_{p\sigma\alpha}(V_{p\alpha} d_{\sigma}^\dag c_{p\sigma\alpha}+\rm{H.c.}).
    \end{aligned}
\end{equation}
where we introduced the dot energy $\ce_d$, $c_{p\sigma\alpha}$ annihilates a fermion in lead $\alpha\in\{R,L\}$ with momentum $p$ and single-particle energy $\ce_{p\alpha}=\ce_p-\mu_\alpha$, possibly biased by a chemical potential $\mu_\alpha$. We notice that the tuning of $\ce_d$ and the possibility of biasing the reservoirs are both within the current experimental capabilities \cite{Esslinger_JPhys,Esslinger_quantized_cond}. As it is usual in impurity problems, the leads can always be considered to be one-dimensional \cite{Hewson,HaugJauho,Giamarchi,GogolinNersesyanTsvelik}. However, in a realistic transport setup they are typically three-dimensional \cite{Esslinger_JPhys,Esslinger_Brantut_transport_science}. The tunneling between the dot and the leads is governed by the amplitudes $V_{p\alpha}$. For most purposes, the leads are fully described by the level width function \cite{HaugJauho} $\Gamma_\alpha(\omega)\equiv 2\pi\sum_{p} \abs{V_{p\alpha}}^2\delta(\omega-\ce_{p})\equiv \Gamma_\alpha \xi(\omega)$, which we parametrize in terms of the energy scale $\Gamma_\alpha$, which contributes to the tunneling rate $\Gamma_T\equiv\sum_\alpha \Gamma_\alpha$ of the dot levels in the absence of dissipation, and the shape function $0\le\xi(\omega)\le1$ that sets the bandwidth $W\gg \abs{\ce_d},\,\Gamma_\alpha$. We are going to assume symmetric leads $\Gamma_\alpha=\Gamma$ and a flat shape function $\xi(\omega)=\theta(W-\abs{\omega})$ ($\theta$ being the Heaviside function). We highlight that by assumption no interactions are present in any portion of the system \footnote{{This simplifying assumption is not particularly restrictive, as interactions in the leads can be tuned via Feshbach resonances \cite{Esslinger_JPhys}. Moreover, we expect our model  to be at least qualitatively valid even for weakly repulsive interactions, in the spirit of Fermi liquid theory (since realistic leads are three-dimensional).}}, so that the only source of correlations are the two-body losses. 
\par We are interested in the regime of strong dissipation, in which the decay rate $\gamma$ is the largest energy scale. We will consider both the dynamics of the system after turning on the dissipation, and the properties of the local steady state that forms around the dot at long times. Since the dissipation removes pairs of opposite-spin fermions, it is clear that in the limit $\gamma\to+\infty$ the system remains confined to a subspace with no double occupancies on the dot site(s). In the case of a single site, this subspace is spanned by three dark states, namely states which are simultaneously eigenstates of $H_\textup{dot}$ and annihilated by the jump $L$: the empty dot $\ket{0}$ and the singly occupied dot $\ket{\sigma}$. The dynamics within the dark subspace is generated by the exchange of particles with the leads, according to the Hamiltonian:
\begin{equation}\label{eq: eff Hamiltonian}
\begin{aligned}
     H_\textup{eff}=&\ce_d\sum_\sigma X_{\sigma\sigma}+\sum_{p\sigma\alpha} \ce_{p\alpha} c_{p\sigma\alpha}^\dag c_{p\sigma\alpha}\\
    &+\sum_{p\sigma\alpha}(V_{p\alpha} X_{\sigma0} c_{p\sigma\alpha}+\mathrm{H.c.})~,
\end{aligned}
\end{equation}
where the operators $X_{\sigma\sigma}\equiv\ketbra{\sigma}$, $X_{\sigma0}=X_{0\sigma}^\dag\equiv\ketbra{\sigma}{0}$ and $X_{00}\equiv\ketbra{0}$ are known as Hubbard operators \cite{Coleman}. The Hamiltonian \eqref{eq: eff Hamiltonian} is the well-known Anderson impurity model (AIM) with infinite repulsion \cite{AndersonImpurityModel,Bickers,Hewson,Coleman}, a strongly correlated model whose properties have been extensively studied. In particular, the model is known to show the Kondo effect---namely, its low-energy physics is dominated by a narrow many-body resonance pinned at the chemical potential of the leads. 
\par It is important to understand to what extent the Kondo phenomenology associated with the AIM survives once we move away from the ideal $\gamma\to+\infty$ limit. As long as $\gamma$ is still the largest energy scale of the system we can derive corrections to the dynamics in the constrained subspace by means of an adiabatic elimination \cite{Garcia-Ripoll_2009} or, equivalently, via a dissipative generalization of the Schrieffer-Wolff transformation \cite{Kessler}, the details of which are reported in \cite{SM}. Physically, in presence of a strong dissipation $\gamma\gg\Gamma$, any double occupancy on the dot site is rapidly removed from the system before it can be replenished from the leads during a transient that lasts a few $\gamma^{-1}$. The dynamics at later times are effectively projected on the subspace of no double occupancies \footnote{This is not strictly true, of course: there is a residual double occupancy of order $\expval*{d_\uparrow^\dag d_\uparrow d_\downarrow^\dag d_\downarrow}\sim\gamma^{-2}$, that is responsible for the $1/\gamma$ suppression of the effective loss rate $\kappa_\sigma$ \cite{SM}---an instance of the well-known Zeno effect \cite{PRL_2BLoss_Zeno}.}, in which it is described by the master equation
\begin{equation}\label{eq: eff me}
    \dv{}{t}\rho(t)=-\ii\comm{H_\textup{eff}}{\rho(t)}+L_\textup{eff}\rho(t)L_\textup{eff}^\dag-\frac{1}{2}\{L_\textup{eff}^\dag L_\textup{eff},\rho(t)\}~,
\end{equation}
where the Hamiltonian \eqref{eq: eff Hamiltonian} governs the coherent part, while there is a residual dissipation with jump operator 
\begin{equation}
    L_\textup{eff}\equiv \frac{2}{\gamma^{1/2}}\sum_{p\sigma\alpha}\sigma V_{p\alpha}X_{0\bar{\sigma}}c_{p\sigma\alpha}~.
\end{equation}
The effective dissipation accounts for the virtual processes in which a fermion with spin $\sigma$ hops from a lead to the dot while the latter is already occupied by a fermion of opposite spin $\bar{\sigma}$, causing both fermions to be lost from the system due to the dissipation. 
\par The effective master equation \eqref{eq: eff me} introduces two new energy scales into the problem: the Kondo temperature  $T_K\equiv (\Gamma_T W/2)^{1/2}\exp[-\pi(\mu-\ce_d)/\Gamma_T]$ \cite{Hewson,Coleman,LangrethNordlander2}, governing the long-time behavior of the unitary dynamics generated by $H_\text{eff}$ \footnote{{With some exceptions, depending on the initial state, as noticed in \cite{PRL_spin_decay1}. The regimes of model \eqref{eq: eff me} analyzed here do not seem to belong to the exceptional cases.}}, and the residual loss rate, that at the leading order in $\Gamma_\alpha/\gamma$ and for a flat width function $\xi(\omega)=\theta(W-\abs{\omega})$ is given by \cite{SM}
\begin{equation}
\kappa_\sigma=2\sum_\alpha\frac{\Gamma_\alpha}{\pi\gamma}(\mu_\alpha+W)~.
\end{equation}
In general, $\kappa_\sigma$ is non-universal in the sense that it depends on the specific band shape $\xi(\omega)$, but can be estimated as $\kappa_\sigma\sim \order{\Gamma_T W/\gamma}$---roughly speaking, it is proportional to the total {density} of fermions in the reservoirs. 
Since the Kondo effect is generated by coherent tunneling processes, we expect the residual dissipation to compete against it, i.e., that a lowering of $\gamma$ (which increases $\kappa_\sigma$) should wash away the Kondo effect. Indeed, in the limit of a vanishing $\gamma$ we recover the noninteracting model \eqref{eq: Hamiltonian}, that features no Kondo effect. 
\par The effective master equation \eqref{eq: eff me} has been derived assuming that $\gamma\gg W,\,\abs{\ce_d},\,\Gamma$. We can further estimate its regime of validity by computing the probability of a double occupancy on the dot. As shown in \cite{SM}, we obtain $\delta\equiv\expval*{d_\uparrow^\dag d_\uparrow d_\downarrow^\dag d_\downarrow}\sim\order{\kappa_\sigma/\gamma}\sim\order{\Gamma_T W/\gamma^2}$. If we impose a somewhat arbitrary threshold $\delta\lesssim 10^{-2}$ for neglecting double occupancies, we obtain that $\gamma\gtrsim 10(\Gamma_T W)^{1/2}$. Since in our calculations we take $\Gamma_T\sim 10^{-2}W$, we can see that $\gamma$ can be almost as low as $W$.
\subsection{Signatures of Kondo}
For a strictly infinite dissipation, the dynamics are governed by the effective AIM Hamiltonian \eqref{eq: eff Hamiltonian}, which guarantees the presence of Kondo physics---such as the Kondo resonance in the impurity spectral function and the maximal differential conductance at zero bias. Therefore, a first important task is to assess to what extent a finite but large value of $\gamma$ alters the well-known Kondo features. A second interesting question is how these features emerge from the uncorrelated system at $\gamma=0$ once the dissipation is increased.
\par During the years, a large number of different numerical techniques have been employed to tackle the dynamics of the AIM and of the Kondo model for isolated systems---to mention only some of the most recent, we can list methods based on the time-dependent variational principle \cite{PRL_variationalKondo,PRB_variationalKondo}, on the influence functional \cite{TimeMPS_LeroseAbanin} and on matrix product states \cite{MPS_AIM_Wauters}. In the present work we need to deal with dissipative dynamics, and we employ two different numerical approaches. One is based on quantum trajectories \cite{DaleyQuantumTrajectories} and on a variational Ansatz for the state of the system along the trajectories \cite{BravyiGosset_complexity,SGS_SnymanFlorens}. The detailed description of this method and its main results is the subject of the companion paper \cite{companion}. The advantage of this method is that it can be applied for any value of $\gamma$, so that we can observe the full crossover towards the Kondo phenomenology as the dissipation is increased. The main observables accessible to the method are currents and single-time averages and correlation functions.
\par In the present work, we take an alternative approach, and work directly with the effective model \eqref{eq: eff Hamiltonian} and the residual dissipation $L_\textup{eff}$ to observe the effects of a large but finite $\gamma$. In this regime, we use the slave boson representation \cite{SlaveBosons, Coleman,slaveBosons_Barnes} and we apply the widely used non-crossing approximation (NCA) \cite{Hewson, Bickers,PRB_Nonequilibrium_NCA,transport_aim,PRB_noneq_aim,LangrethNordlander1,LangrethNordlander2} to derive Kadanoff-Baym equations for the Keldysh Green's functions of the auxiliary particles \cite{SM}. To the same leading perturbative order as the NCA self-energy, it suffices to treat the effective dissipation at the mean-field level. The crucial feature of this extended NCA is that the resulting approximation is conserving \cite{BaymConserving,LangrethNordlander1,LangrethNordlander2,Fabrizio, PRB_Nonequilibrium_NCA,transport_aim,PRB_noneq_aim}---namely, the approximate dynamics respect the relevant conservation laws---a necessary feature for real-time simulations. The NCA is known to have a few shortcomings, such as violating the Fermi liquid relations at zero temperature and introducing spurious features in some impurity regimes \cite{MullerHartmann,ComparisonNCA-NRG,PhysRevB.70.165102,KirchnerKroha}. However, it is the simplest, conserving diagrammatic method that is able to reveal the emergence of the Kondo resonance in the relevant regime $\ce_d<\mu$, and it is capable of reproducing observables like the conductance through the dot with good accuracy. Since we are more interested in finding typical (qualitative) signatures of Kondo physics rather than providing a quantitatively accurate analysis of a realistic model, we shall not attempt to overcome the limitations on the NCA in the present work.
\par Before describing the main results, we wish to remark that our dissipative model leads to an effective AIM physics only in the sense that a local steady state forms around the dot, as it is common with dissipative impurities \cite{Dolgirev,PRB_Alba_Carollo,tonielli2019orthogonality,tonielli2020ramsey,FromlPRB1,FromlPRB2,FromlPRL,chaudhari2022zeno,javed2023counting,javed2024zeno,UniversalNoneq_DoyonAndrei,SchiroImpurity,SchiroMC}. Mathematically, the local steady state corresponds to the limit $\lim_{t\to\infty}\lim_{\Omega\to\infty}\rho(t)$, where $\Omega$ is the volume of the leads. The two limits do not commute: For finite leads, the system will eventually reach the true stationary manifold, which is spanned by Dicke states \cite{Rey_Hot_reactive_fermions,SM}, including the vacuum state. In the present context, the Dicke states are best understood as eigenstates of the quadratic Hamiltonian $H$ that have no double occupancies (neither in the dot, nor in the leads) and thus are not affected by the losses \footnote{By the same token, they are exact (excited) eigenstates of the usual AIM, too.}. These states have correlations that are distinct from those associated with Kondo physics, and which Dicke states are reached at the end of the dynamics is determined by the initial (conserved) value of the spin. We will further discuss them in the last section of this paper and in \cite{SM}. 
A related yet possible scenario is the formation of a finite size ferromagnetic bubble around the impurity site. The larger the bubble, the stronger the suppression of losses. On the other hand, larger bubbles are energetically unfavorable and their formation requires a less likely statistical fluctuation. In the NCA analysis the possibility of ferromagnetic bubbles is not included, because it requires introducing spin symmetry breaking. We will discuss conditions for observing ferromagnetic bubbles in the companion paper \cite{companion}.
\par We have computed the real-time dynamics of the effective model after a quench of the tunneling $\Gamma_T$, and we have analyzed its properties in the local steady state at late times. We first discuss the stationary properties, which are clearer to understand. The main signature of Kondo behavior is the presence of the Kondo resonance in the impurity spectral function $A_\sigma(\omega)\equiv-\Im G^R_{d\sigma}(\omega)/\pi$, where $G^R_{d\sigma}(\omega)$ is the Fourier transform of the retarded impurity Green's function $G^R_{d\sigma}(t-t^\prime)\equiv -\ii \theta(t-t^\prime)\expval{\{d_{\sigma}(t),d^\dag_\sigma(t^\prime)\}}$ \cite{SM}, computed in the local steady state \footnote{See \cite{SM} for the full time-dependent generalization of $A_\sigma(\omega)$, displayed in the Figure.}. We show it in the central panel of \fig~\ref{fig: aim} for an impurity tuned to the Kondo regime $\ce_d=-2\Gamma_T$ in presence of unbiased reservoirs. We observe the typical two-peak shape that one finds in the AIM \cite{Hewson,Coleman,Bickers}. There is a broad peak centered close to the single-particle energy $\ce_d$, whose width is set by the original tunneling rate $\Gamma_T$. This peak reflects the rapid exchange of charge between the dot and the leads. The second peak is the Kondo resonance: a much narrower feature, pinned at the reservoirs' chemical potential $\mu_{R,L}=0$, which signals the presence of a long-lived spin degree of freedom. In the $\gamma\to\infty$ limit, the width of the peak is set by the Kondo temperature $T_K$ \cite{Hewson}. In the full model there would be also an extremely broad peak at the same energy $\omega\approx \ce_d$, with width of order $\gamma$ and a highly reduced height, corresponding to the doubly-occupied dot. The effect of a finite dissipation is to suppress and broaden the Kondo peak as a function of the ratio $\kappa_\sigma/T_K$, where $\kappa_\sigma$ is the Zeno-suppressed effective rate of particle loss from the system \cite{SM}. 
\par We can have an intuition on why a finite dissipation works against the formation of the Kondo peak by noticing \cite{transport_aim} that it reflects the overlap between states of the system differing by one fermion (thus separated in energy by the chemical potential of the leads)---two ground states, for the Hamiltonian case (i.e., $\gamma\to+\infty$). With a finite effective dissipation, the system lacks a ground state and all eigenstates (in the projected subspace) acquire a lifetime of order $\kappa_\sigma^{-1}$, leading to the broadening of the spectral features. 
As we can observe in the central panel of \fig~\ref{fig: aim}, this blurring occurs in a continuous fashion as a function of $\gamma^{-1}$. The Kondo resonance becomes significantly smeared out for $\kappa_\sigma$ grater than a few times $T_K$. 
\par While the shape of the spectral function is a clear theoretical signature of the Kondo effect, its measurement through radio-frequency spectroscopy can be challenging in experiments with ultracold atoms, since it requires a sharp frequency resolution to reveal the Kondo peak. As mentioned in the introduction, a more accessible probe is provided by transport. The existence of the Kondo resonance can be assessed directly by measuring the differential conductance through the dot \cite{HaugJauho,transport_aim,transport_aim,PRB_noneq_aim,PRB_Nonequilibrium_NCA}. The latter is defined as $g(\Delta\mu,t)\equiv \dv*{I(\Delta\mu,t)}{\Delta\mu}$, where $I(\Delta\mu,t)=-2^{-1}\dv*{(N_L(t)-N_R(t))}{t}$ is the current flowing from the left reservoir to the right one for a given chemical potential bias $\Delta\mu$ [with $N_\alpha(t)\equiv\sum_{p\sigma}\expval*{c_{p\sigma\alpha}^\dag c_{p\sigma\alpha}}(t)$ the number of particles in reservoir $\alpha$]. This definition, which is the one employed in closed systems and also in experiments involving dissipation \cite{Esslinger_PRL, Esslinger_PRA_theory,Esslinger_PRL_23,Ott_PRL_bistability,PRL_lattice_transport_local_loss,PRA_Uchino,PRRes_Esslinger}, does not distinguish between particles leaving a reservoir because of transport or because of losses \footnote{In principle, one could describe the particle losses as caused by the coupling to a third, empty reservoir \cite{PRA_Uchino}. This point of view might be sometimes useful for interpreting the results.}. At zero temperature, the zero-bias (i.e., linear response) conductance directly measures the height of the Kondo peak \cite{HaugJauho,transport_aim}: 
\begin{equation}\label{eq: conductance}
\begin{aligned}
    g(0,t)=
    &g_{MW}(0,t)+\frac{\Gamma(\mu)}{\pi \gamma}n_d(t)~,
\end{aligned}
\end{equation}
where $n_d(t)\equiv\sum_\sigma\expval*{d_\sigma^\dag d_\sigma}(t)$ is the dot population at time $t$. See \cite{SM} for the full expression for $g_{MW}(0,t)$. The first term of \eq~\eqref{eq: conductance} converges at later times to the usual Meir-Wingreen formula \cite{MeirWingreenFormula,HaugJauho} $g_\infty(0)=2^{-1}\Gamma(\mu) \sum_\sigma A_\sigma(\mu)$. This term provides the dominant contribution, and---although formally equivalent to the $\gamma\to+\infty$ expression---it already includes most of the effects of the residual dissipation on the coherent part of transport. The second term is a small correction related to the effective losses, and it is positive, because the most populated lead at a higher chemical potential suffers an effective loss rate that is higher than the other, yielding a net current in the same direction as transport. Since we observed that a finite dissipation rate $\gamma$ decreases the height of the Kondo resonance, we expect that the zero-bias conductance $g(0,t)$ will be decreased as well. This behavior can be seen in the inset of \fig~\ref{fig: aim} for the stationary state---the conductance decreases rapidly with $\gamma^{-1}$, from a value close to the maximal one \footnote{The curves at lower values of $\ce_d$  are rather far from the maximal conductance, because for them $T_K\ll\kappa_\sigma$ even with the largest value of $\gamma$ considered in the inset, so that the Kondo resonance is suppressed. We also notice that the various curves should not converge for $\gamma\to\infty$, since in the AIM with infinite repulsion $g_\infty(0)<2/h$ and depends on $\ce_d$ \cite{transport_aim}.}, $2/h$ (with $h$ Planck's constant) \cite{transport_aim}. 
\par As a final signature of the Kondo effect, which may prove suitable to experiments, in the right panel of \fig~\ref{fig: aim} we show the relaxation rate of the impurity magnetization $\expval{\sigma^z}(t)\equiv \sum_\sigma \sigma\expval*{d_\sigma^\dag d_\sigma}(t)$, for an initially polarized state $\ket{\psi_0}=\ket{\uparrow}_d\ket{FS}_l$, where $\ket{FS}_l$ is the ground state of the unbiased leads. We observe that, when the dot is in the Kondo regime $\ce_d\lesssim -\Gamma_T$, and after an initial transient, the magnetization decays exponentially $\expval{\sigma^z}(t)\sim \ee^{-\Gamma_\text{decay} t}$, and that the decay rate is strongly suppressed as $\gamma$ increases. Indeed, in the limit $\gamma\to+\infty$, $\Gamma_\text{decay}$ is expected to be proportional to $T_K$ \cite{MPS_AIM_Wauters,PRL_spin_decay1,PRB_spin_decay1}, as we confirm in the Figure by showing its exponential dependence on $\ce_d/\Gamma_T$. For decreasing dissipation rate $\gamma$, $\Gamma_\text{decay}$ converges to $\kappa_\sigma$, which is independent of $\ce_d$ (but still much smaller than $\Gamma_T$). 
\subsection{Higher spin models} 
\begin{figure}
    \centering
    \includegraphics[width=.9\linewidth]{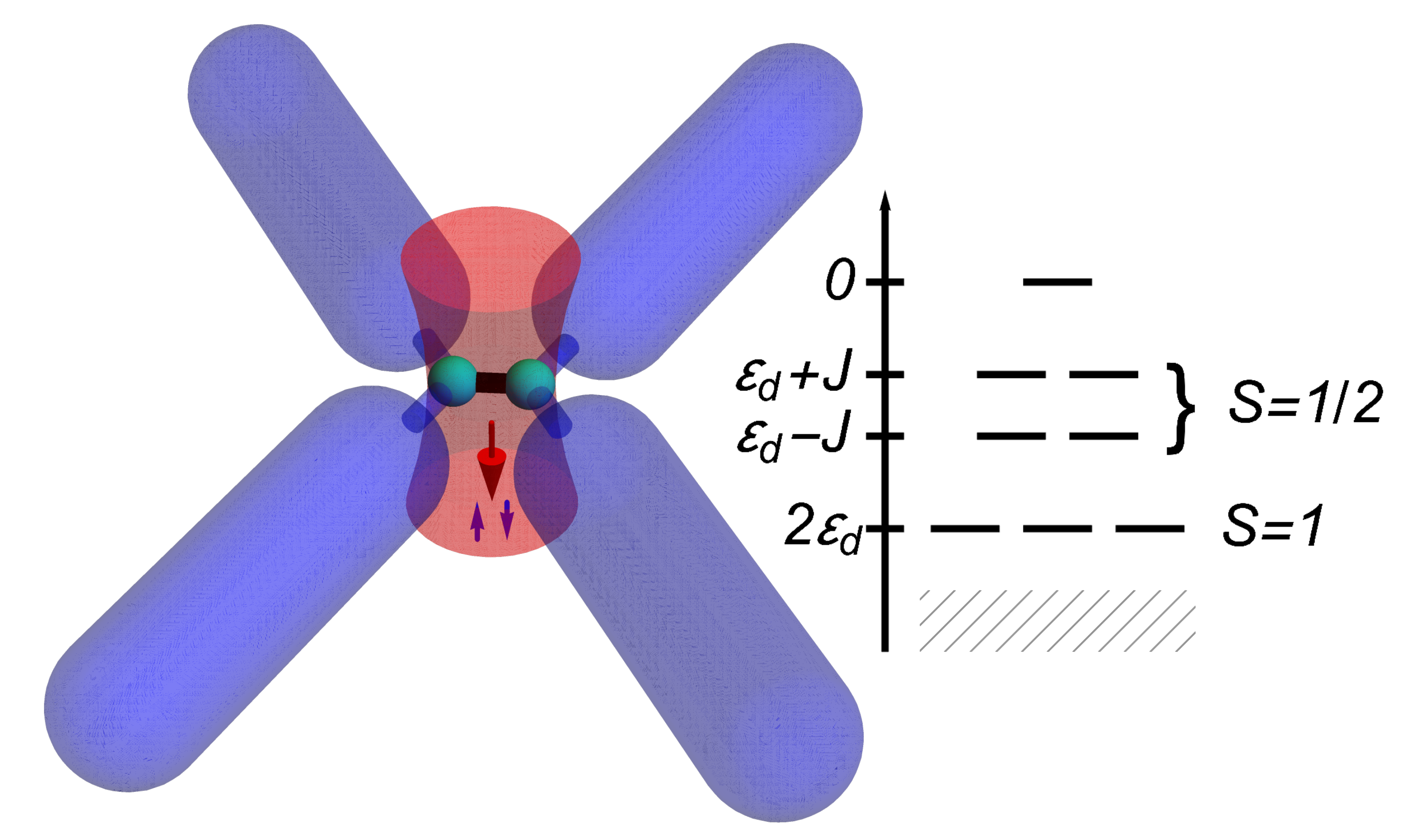}
    \caption{Realizing generalized Kondo models with dissipation. Left panel: Graphical representation of the setup in the case of two dissipative sites coupled to four leads. Right panel: energies of the dark states of the isolated dot with two sites. The hatched area represents the dissipated levels.}
    \label{fig: spin 1}
\end{figure}
We have shown how local two-body losses can be used to realize a prototypical strongly correlated model with ultracold atoms, and how the resulting Kondo physics competes with the residual dissipation. We can obtain more exotic Hamiltonians if the dissipation involves more than one site, as depicted in the left panel of \fig~\ref{fig: spin 1} for the case of two sites and four leads.  
For concreteness, we take the dot to be a linear chain of of $\ell_d$ sites with nearest-neighbor hopping $J$ and onsite energies $\ce_d$, with open boundary conditions. 
\par As in the case of a single dot site, we want to confine the dynamics of the system to the dark subspace of the dot, so that a weak coupling to the leads will induce only a small residual dissipation. More details on the construction can be found in \cite{companion}. The right panel of \fig~\ref{fig: spin 1} depicts the dark subspace for two impurity sites. The (strong) rotational invariance of the Lindblad equation for the isolated dot sites allows to build the dark states as simply the Dicke states \cite{Dicke, PRA_atomic_coh_states, Rey_Hot_reactive_fermions} associated with $H_\textup{dot}$ \cite{SM}.
The resulting dark subspace is organized in multiplets of increasing spin $S=0,\dots \ell_d/2$ and particle numbers $n=2S$, with higher spins possessing lower energy (although there are $\binom{\ell_d}{2S}$ different multiplets with the same spin $0<S<\ell_d/2$ but different energies, where $\binom{\cdot}{\cdot}$ is the binomial coefficient). We notice that this procedure is completely general, in the sense that it does not rely on any particular structure of the dot region, except for spin rotational invariance. In this regard, our treatment is similar to that of the Hubbard Hamiltonian with infinite repulsion \cite{PRL_infinite_U,PRB_infinite_U,JStat_infinite_U}.
\par In the setup with two sites there is an extra complication with respect to the single-site case: the slowest-decaying eigenstate of the Lindbladian has a decay rate $\gamma_1$ which is nonmonotonic in $\gamma$, with a maximum $\gamma_1^\ast$ at $\gamma=8J$ followed by a slow decrease $\gamma_1\sim J^2/\gamma$. In contrast, all other bright states have decay rates of order of $\gamma$. So, to keep the occupation of all bright states suppressed by dissipation, we need to have a large hopping $J\sim\gamma$, and hence a large dot energy $\ce_d\sim\gamma$ to maintain the spin $1$ states at energy lower than the spin $1/2$ ones, $2\ce_d<\ce_d-J$. The nonmonotonic behavior of the smallest decay rate, and the consequent need for fine tuning, is likely to be generic for higher $\ell_d$ \cite{SM}.
\par If we introduce back the coupling to the leads, the latter will mediate transitions between multiplets with neighboring values of $S$, and we obtain a model that belongs to the family of the ionic models \cite{Hewson,Bickers,Hirst_ionic_model}, that have been extensively studied in the context of magnetic impurities in metals. These models can be mapped on higher-spin Kondo models by the usual Schrieffer-Wolff transformation \cite{SchriefferWolff,Hewson,Coleman}. In the case of two sites, the lowest-energy multiplet would act like a spin $1$ impurity. 
\par While two-body losses lead naturally to higher-spin generalizations of the Kondo model, the systems considered in this work are all single-channel, in the sense that the particle exchange between the reservoirs makes them behave like a single one at thermodynamic equilibrium \cite{GogolinNersesyanTsvelik,Giamarchi,Hewson,Cox_Zawadowski_exotic_Kondo}. Nevertheless, the experimental control over the geometry and couplings of the ultra-cold atomic setups may be employed to achieve the necessary conditions for a truly multi-channel Kondo model, following the various approaches present in the literature \cite{PRB_Kiselev_two_level_Kondo, PRB_Kiselev_multistage_Kondo,PhysRevLett.90.136602,Nature_2ck}.



\begin{acknowledgments}
    We acknowledge useful discussions with D. Sels, A. Gomez Salvador, R. Andrei, M. Kiselev. M.S. and J.M. have been supported by the DFG through the grant HADEQUAM-MA7003/3-1. Y.Q. and E.D. acknowledge support by the SNSF project 200021\_212899, the NCCR SPIN, a National Centre of Competence in Research, funded by the Swiss National Science Foundation (grant number 225153), the Swiss State Secretariat for Education, Research and Innovation (contract number UeM019-1). E.D. also acknowledges support from the ARO grant number W911NF-20-1-0163. T.E., Y.Q., and E.D. acknowledge funding by the ETH grant. J.M. acknowledges the Pauli Center for hospitality. We gratefully acknowledge the computing time granted through the project “DysQCorr” on the Mogon II supercomputer of the Johannes Gutenberg University Mainz (\url{hpc.uni-mainz.de}), which is a member of the AHRP (Alliance for High Performance Computing in Rhineland Palatinate, \url{www.ahrp.info}), and the Gauss Alliance e.V. 
\end{acknowledgments}

\bibliography{biblioDissKondo.bib}
\end{document}


\title{Supplemental Material --- Dissipative realization of Kondo models}

\author{Martino Stefanini}
\affiliation{Institut f\"ur Physik, Johannes Gutenberg-Universit\"at Mainz, D-55099 Mainz, Germany}
\author{Yi-Fan Qu}
\affiliation{Institute for Theoretical Physics, ETH Zurich, 8093 Zurich, Switzerland}
\author{Tilman Esslinger}
\affiliation{Institute for Quantum Electronics \& Quantum Center, ETH Zurich, 8093 Zurich, Switzerland}
\author{Sarang Gopalakrishnan}
\affiliation{Department of Electrical and Computer Engineering, Princeton University, Princeton, NJ 08544}
\author{Eugene Demler}
\affiliation{Institute for Theoretical Physics, ETH Zurich, 8093 Zurich, Switzerland}
\author{Jamir Marino}
\affiliation{Institut f\"ur Physik, Johannes Gutenberg-Universit\"at Mainz, D-55099 Mainz, Germany}

\date{\today}

\maketitle
\section{Derivation of the effective dynamics}
In this Section we derive the effective master equation governing the dynamics of the single-site dissipative dot at large dissipation.
\par We consider the strongly dissipative limit in which the loss rate is the largest energy scale $\gamma\gg W,\, \abs{\ce_d},\,\Gamma$. In other words, we will take the purely dissipative dynamics as the zeroth-order solution and expand in the Hamiltonian contribution to the Lindblad equation. This approach is slightly different from the usual Schrieffer-Wolff mapping of the Anderson impurity model to the Kondo model \cite{Coleman,Hewson}, which treats the uncoupled dot-leads system as the unperturbed dynamics and expands in $H_\textup{tun}$, i.e. in $\Gamma$. In the present model, this would amount to $\gamma \sim W \sim \abs{\ce_d}\gg\Gamma$. Although it is possible to adopt this approach also in the dissipative case, it yields considerably more complicated expressions for the effective dynamics, while adding very little to the physical description of the system.
\par The starting point of our analysis is the purely dissipative dot site, decoupled from the rest of the system. The master equation becomes
\begin{equation}
    \dv{}{t}\rho=\mathcal{L}_0\rho~,
\end{equation}
where
\begin{equation}
    \mathcal{L}_0\rho\equiv \gamma\big(L\rho L^\dag-\tfrac{1}{2}\{L^\dag L,\rho\}\big)
\end{equation}
We label the states of the dissipative site as $\ket{\alpha}$, where $\alpha\in\{0,\uparrow,\downarrow,d=\uparrow\downarrow\}$, and vectorize the density matrix as \cite{Garcia-Ripoll_2009} $\rho=\sum_{\alpha\beta}\rho_{\alpha\beta}\ketbra{\alpha}{\beta}\to \vmr{\rho}=\sum_{\alpha\beta}\rho_{\alpha\beta}\vmr{\alpha\beta}$. Then, it is easy to see that in the 16-dimensional $\vmr{\alpha\beta}$ basis the Liouville superoperator $\mathcal{L}_0$ is almost diagonal:
   \begin{align*}
       &\mathcal{L}_0\vmr{\alpha\beta}=0\quad\text{for all }\alpha,\,\beta\in\{0,\uparrow,\downarrow\}~,\\
       &\mathcal{L}_0\vmr{\alpha d}=-\frac{\gamma}{2}\vmr{\alpha d}\quad\text{for all }\alpha\in\{0,\uparrow,\downarrow\}~,\\
       &\mathcal{L}_0\vmr{d\alpha}=-\frac{\gamma}{2}\vmr{d\alpha}\quad\text{for all }\alpha\in\{0,\uparrow,\downarrow\}~,\\
       &\mathcal{L}_0\vmr{dd}=\gamma(\vmr{00}-\vmr{dd})
   \end{align*}
The only coupled states are $\vmr{00}$ and $\vmr{dd}$, and the diagonalization of the corresponding $2\times2$ matrix yields the remaining two eigenvalues of $\mathcal{L}_0$:
\begin{subequations}
    \begin{align}
    &\left.
    \begin{aligned}
        &\mathcal{L}_0\vmr{\phi^R_0}=0,\quad\vmr{\phi^R_0}=\vmr{00}\\
        &\vml{\phi^L_0}\mathcal{L}_0=0,\quad\vml{\phi^L_0}=\vml{00}+\vml{dd}\\
    \end{aligned}
    \right\}\quad \lambda_0=0\\
    &\left.
    \begin{aligned}
        &\mathcal{L}_0\vmr{\phi^R_d}=-\gamma \vmr{\phi^R_d},\quad\vmr{\phi^R_d}=-\vmr{00}+\vmr{dd}\\
        &\vml{\phi^L_d}\mathcal{L}_0=-\gamma \vml{\phi^L_d},\quad\vml{\phi^L_d}=\vml{dd}\\
    \end{aligned}
    \right\}\quad \lambda_2=-\gamma
    \end{align}
\end{subequations}
As usual, we are normalizing the states as $(\phi_a^L\vert\phi_b^R)=\delta_{ab}$. Summing up, there are nine stationary (dark) states with eigenvalue $\lambda_0=0$, six states with eigenvalue $\lambda_1=-\frac{\gamma}{2}$, and a non-degenerate state with $\lambda_2=-\gamma$. For later use, we report here the superprojectors on the three eigenspaces:
\begin{subequations}
    \begin{align}
        &\mathcal{P}_0\equiv\vmr{00}\vml{dd}+\sum_{\alpha,\beta<d}\vmr{\alpha\beta}\vml{\alpha\beta}~,\\
        &\mathcal{P}_1\equiv\sum_{\alpha<d}(\vmr{\alpha d}\vml{\alpha d}+\vmr{d\alpha}\vml{d\alpha})~,\\
        &\mathcal{P}_2\equiv \vmr{dd}\vml{dd}-\vmr{00}\vml{dd}
    \end{align}
\end{subequations}
The $\mathcal{P}_0$ projector defines the slow subspace in which the dynamics will be confined at large dissipation. It is useful to express the above superprojectors in terms of operators in the ordinary Hilbert space:
\begin{subequations}\label{eq: projectors}
    \begin{align}
        &\mathcal{P}_0\rho=X_{0d}\,\rho\,X_{d0}+X_<\,\rho\, X_<~,\\
        &\mathcal{P}_{1}\rho=X_<\,\rho\, X_d+X_d\,\rho\,X_<~,\\
        &\mathcal{P}_2\rho=X_d\,\rho\, X_d-X_{0d}\,\rho\,X_{d0}~,
    \end{align}
\end{subequations}
where we have introduced the ordinary projectors (Hubbard operators \cite{Hewson,Bickers,Coleman})
\begin{subequations}
    \begin{align}
        &X_{\alpha\beta}\equiv\op{\alpha}{\beta}~,\\
        &X_\alpha\equiv X_{\alpha\alpha}=\dyad{\alpha}~,\\
        & X_<\equiv\sum_{\alpha<d} X_\alpha
    \end{align}
\end{subequations}
We turn to consider the full Lindblad master equation
\begin{equation}
    \dv{}{t}\rho=(\mathcal{L}_0+\mathcal{L}_H)\rho~,
\end{equation}
where $\mathcal{L}_H=-\ii\comm{H}{\cdot}$ is the Hamiltonian part, which we will treat as a perturbation. Following \cite{GeneralizedSchriefferWolff} and \cite{Garcia-Ripoll_2009}, the effective dynamics in the slow subspace $\mathcal{P}_0\rho=\rho$ in the $\gamma\to+\infty$ limit is generated by the effective Lindblad superoperator $\mathcal{L}_\textup{eff}=\mathcal{L}_1+\mathcal{L}_2+\order{\gamma^{-2}}$, where
\begin{subequations}
\begin{align}
    &\mathcal{L}_1=\mathcal{P}_0 \mathcal{L}_H \mathcal{P}_0~,\\
    &\mathcal{L}_2=-\mathcal{P}_0 \mathcal{L}_H (\mathcal{Q}_0 \mathcal{L}_0 \mathcal{Q}_0)^{-1} \mathcal{L}_H\mathcal{P}_0~,\label{eq: L2}
\end{align}
\end{subequations}
and $\mathcal{Q}_0=\one-\mathcal{P}_0$ is the complementary projector to $\mathcal{P}_0$.
The first-order term is just the Hamiltonian part projected on the stationary subspace of $\mathcal{L}_0$, $ \mathcal{L}_1\rho=\mathcal{P}_0 \mathcal{L}_H \mathcal{P}_0\rho$. We classify the Hamiltonian terms according to the commutativity of the corresponding superoperator $-\ii\comm{H}{\cdot}$ with $\mathcal{P}_0$. The dot Hamiltonian $H_\textup{dot}$ and the leads' one $H_\textup{leads}$ both commute (the latter since $\mathcal{P}_0$ acts as the identity on the leads states). The tunneling term $H_\textup{tun}$ is the only one which does not commute, since it connects the slow subspace $\mathcal{P}_0\rho=\rho$ with the other subspaces. The decomposition $d_\sigma=X_{0\sigma}+\sigma X_{\bar{\sigma}d}$ induces the further subdivision $H_\textup{tun}=H_\textup{tun}^0+H_\textup{tun}^1$, with
\begin{equation}
    \begin{aligned}
        &H_\textup{tun}^0=\sum_\sigma(X_{\sigma0} \Psi_\sigma+\Psi_\sigma^\dag X_{0\sigma})~,\\
        &H_\textup{tun}^1=\sum_\sigma\sigma(X_{d \bar{\sigma}} \Psi_\sigma + \Psi_\sigma^\dag X_{\bar{\sigma} d})~,
    \end{aligned}
\end{equation}
where $\Psi_\sigma\equiv\sum_{p\alpha} V_{p\alpha} c_{p\sigma\alpha}$. Notice that the projectors $X_{\sigma0}$, $X_{d\sigma}$ and their conjugates are of fermionic nature in the sense that they anticommute with $\Psi_\sigma$, $\Psi_\sigma^\dag$, while all other projectors are bosonic (i.e. commuting with the $\Psi$s). The first term $H_\textup{tun}^0$ commutes with $\mathcal{P}_0$, while the second is annihilated by it. So, if we call $H_0=H_\textup{dot}+H_\textup{leads}+H_\textup{tun}^0$ the part of the Hamiltonian that commutes with $\mathcal{P}_0$, we have
\begin{equation}
    \mathcal{L}_1\rho=\mathcal{P}_0 (\mathcal{L}_{H_0}+\mathcal{L}_{H_\textup{tun}^1}) \mathcal{P}_0\rho=\mathcal{L}_{H_0}\mathcal{P}_0^2\rho +\mathcal{P}_0 \mathcal{L}_{H_\textup{tun}^1}\mathcal{P}_0\rho=\mathcal{L}_{H_0}\rho
\end{equation}
where we used $\rho =\mathcal{P}_0\rho$. The Hamiltonian $H_0$ is the Anderson impurity model with infinite dissipation quoted in \eq~(1) in the main text.
\par The second-order term $\mathcal{L}_2$ yields the effective dissipation. Since $\mathcal{Q}_0 \mathcal{L}_0 \mathcal{Q}_0=-\gamma/2 \mathcal{P}_1-\gamma\mathcal{P}_2$, we have $(\mathcal{Q}_0 \mathcal{L}_0 \mathcal{Q}_0)^{-1}=-2/\gamma\, \mathcal{P}_1-1/\gamma\,\mathcal{P}_2$. We evaluate \eq~\eqref{eq: L2} starting from its rightmost terms:
\begin{equation}
        (\mathcal{Q}_0 \mathcal{L}_0 \mathcal{Q}_0)^{-1} \mathcal{L}_H\mathcal{P}_0\rho=\Big(-\frac{2}{\gamma}\mathcal{P}_1-\frac{1}{\gamma}\mathcal{P}_2\Big)(\mathcal{L}_{H_0}+\mathcal{L}_{H_\textup{tun}^1})\rho=\frac{2}{\gamma}\mathcal{P}_1\mathcal{L}_{H_\textup{tun}^1}\rho~.
\end{equation}
In the expression above, the terms proportional to $\mathcal{L}_{H_0}\rho$ vanish because of the commutativity of $\mathcal{L}_{H_0}$ with both projectors $\mathcal{P}_{1,2}$, which entails
$\mathcal{P}_{1,2}\mathcal{L}_{H_0}\mathcal{P}_{0}=\mathcal{P}_{1,2}\mathcal{P}_{0}\mathcal{L}_{H_0}=0$. The term proportional to $\mathcal{P}_2\mathcal{L}_{H_\textup{tun}^1}\rho$ also vanishes because if $\rho$ belongs to the slow subspace $\mathcal{P}_0$, $\mathcal{L}_{H_\textup{tun}^1}\rho$ yields terms belonging to the $\mathcal{P}_1$ subspace, but not to the $\mathcal{P}_2$ one. Then we expand
\begin{equation*}
    \begin{aligned}
        \frac{2}{\gamma}\mathcal{P}_1\mathcal{L}_{H_\textup{tun}^1}\rho&=\frac{2\ii}{\gamma}\big(X_< H_\textup{tun}^1 \rho X_d-X_< \rho H_\textup{tun}^1 X_d+X_d H_\textup{tun}^1 \rho X_<-X_d \rho H_\textup{tun}^1 X_<\big)\\
        &=\frac{2\ii}{\gamma}\sum_\sigma \sigma \big[\Psi_\sigma^\dag X_{\bar{\sigma}d} \rho X_d-X_< \rho \Psi_\sigma^\dag X_{\bar{\sigma}d}+X_{d\bar{\sigma}} \Psi_\sigma \rho X_<-X_d \rho X_{d\bar{\sigma}}\Psi_\sigma\big]\\
        &=\frac{2\ii}{\gamma}\sum_\sigma \sigma \big[-X_< \rho \Psi_\sigma^\dag X_{\bar{\sigma}d}+X_{d\bar{\sigma}} \Psi_\sigma \rho X_<\big]~,
    \end{aligned}
\end{equation*}
where in the last equality we used $X_{\alpha d}\rho X_{d\beta}=0$ for $\rho$ in the slow subspace. Then,
\begin{equation*}
    \begin{aligned}
        \mathcal{L}_2&=-\mathcal{P}_0\big(\mathcal{L}_{H_0}+\mathcal{L}_{H_\textup{tun}^1}\big)\mathcal{P}_1\mathcal{L}_{H_\textup{tun}^1}\rho=-\frac{2}{\gamma}\mathcal{L}_{H_0}\mathcal{P}_0\mathcal{P}_1\mathcal{L}_{H_\textup{tun}^1}-\frac{2}{\gamma}\mathcal{P}_0\mathcal{L}_{H_\textup{tun}^1}\mathcal{P}_1\mathcal{L}_{H_\textup{tun}^1}\rho=-\frac{2}{\gamma}\mathcal{P}_0\mathcal{L}_{H_\textup{tun}^1}\mathcal{P}_1\mathcal{L}_{H_\textup{tun}^1}\rho\\
        &=-\frac{2}{\gamma}\mathcal{P}_0\sum_{\sigma\tau}\sigma\tau\big(-2 X_{d\bar{\tau}}\Psi_\tau\rho\Psi^\dag_\sigma X_{\bar{\sigma}d}+X_<\rho\Psi_\sigma^\dag X_{\bar{\sigma}\bar{\tau}}\Psi_\tau+\Psi_\sigma^\dag X_{\bar{\sigma}\bar{\tau}}\Psi_\tau\rho X_<\big)\\
        &=\frac{4}{\gamma}\sum_{\sigma\tau}\sigma\tau\big( X_{d\bar{\tau}}\Psi_\tau\rho\Psi^\dag_\sigma X_{\bar{\sigma}d}-\frac{1}{2}\rho\Psi_\sigma^\dag X_{\bar{\sigma}\bar{\tau}}\Psi_\tau-\frac{1}{2}\Psi_\sigma^\dag X_{\bar{\sigma}\bar{\tau}}\Psi_\tau\rho\big)~,
    \end{aligned}
\end{equation*}
where we recognize the dissipative part of the effective Lindblad superoperator,
\begin{equation}
    \mathcal{L}_\textup{eff}\rho=L_\textup{eff}\rho L_\textup{eff}^\dag-\tfrac{1}{2}\{L_\textup{eff}^\dag L_\textup{eff},\rho\}
\end{equation}
with
\begin{equation}\label{eq:lind}
    L_\textup{eff}\equiv\frac{2}{\gamma^{1/2}}\sum_\sigma \sigma X_{0\bar{\sigma}}\Psi_\sigma=\frac{2}{\gamma^{1/2}}\sum_{p\alpha\sigma} \sigma V_{p\alpha}X_{0\bar{\sigma}}c_{p\sigma\alpha}~,
\end{equation}
which is the one quoted in the main text.
\section{Details on the non-crossing approximation}
\begin{figure}
    \centering
    \includegraphics[width=.5\linewidth]{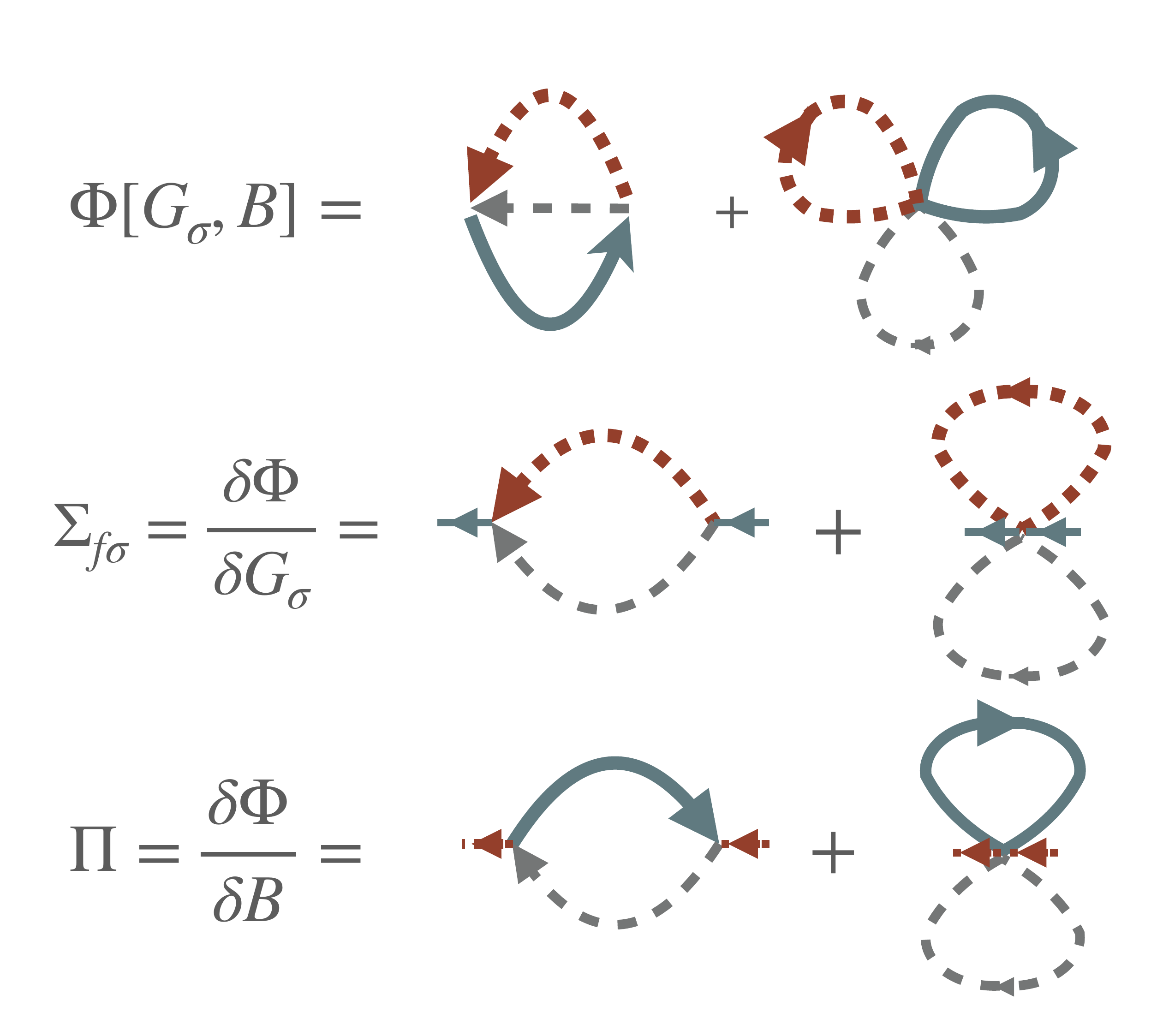}
    \caption{Upper row: diagrammatic representation of the Luttinger-Ward functional truncated to the second order in the dot-leads hopping. The thick, aquamarine lines are the Green's functions of the auxiliary fermions, the dotted red lines are those of the auxiliary bosons and the dashed gray lines are the local ones of the leads. The lower rows of the figures depict the noninteracting fermionic and bosonic self-energies derived from $\Phi$. In each self-energy, the first diagram is the usual NCA self-energy coming from the effective AIM Hamiltonian, while the second term is the mean-field contribution from the effective dissipation. The figures represent the diagrams before the projection onto the physical $Q=1$ subspace.}
    \label{fig: diagrams}
\end{figure}
In this Section, we provide analytical details of our implementation of the non-crossing approximation and on our real-time calculations.
\par The Anderson impurity model with infinite repulsion that represents the effective unitary dynamics of our dissipative model is not suitable to the usual perturbative techniques based on Green's functions, since the Hubbard operators $X_{\alpha\beta}$ do not obey a standard fermionic or bosonic algebra and so their correlation functions cannot be extracted through Wick's theorem. So, we convert the effective Hamiltonian to a form that is suitable to standard perturbative treatments. This is achieved through the slave boson mapping of the Hubbard operators to a boson $b$ and a spinful auxiliary fermion $f_\sigma$ \cite{Coleman, slaveBosons_Barnes, SlaveBosons}: 
\begin{equation}\label{eq: slave bosons}
    \begin{aligned}
        X_{\sigma 0}&=\ketbra{\sigma}{0}=d_{\sigma}^\dag (\one-n_{\bar{\sigma}})=f_\sigma^\dag b\\
        X_{0\sigma}&=\ketbra{0}{\sigma}=(\one-n_{\bar{\sigma}})d_{\sigma}=b^\dag f_\sigma\\
        X_{00}&=\ketbra{0}{0}=(\one-n_{\uparrow})(\one-n_{\downarrow})=b^\dag b
    \end{aligned}
\end{equation}
with the physical subspace defined by $Q\equiv b^\dag b+\sum_\sigma f_\sigma^\dag f_\sigma=1$. The physical meaning of the mapping is that the auxiliary fermions create single occupancies $\ket{\sigma}=f_\sigma^\dag\ket{\text{vac}}$, while the bosons create the empty site $\ket{0}=b^\dag \ket{\text{vac}}$, where $\ket{\text{vac}}$ is the vacuum in the extended Hilbert space. The effective Hamiltonian becomes
\begin{equation}
    H_\textup{eff}=\sum_\sigma\ce_\sigma f_\sigma^\dag f_\sigma+\sum_{p\sigma} \ce_p c_{p\sigma\alpha}^\dag c_{p\sigma\alpha}+\sum_{p\sigma}V_p (b f_\sigma^\dag c_{p\sigma\alpha}+\mathrm{H.c.})~,
\end{equation}
and the effective jump operator reads
\begin{equation}
    L_\textup{eff}=\frac{2}{\gamma^{1/2}}b^\dag\sum_{p\sigma}\sigma V_p f_{\bar{\sigma}}c_{p\sigma\alpha}~,
\end{equation}
corresponding to the Lindbladian dissipative action \cite{Sieberer_2016}:
\begin{equation}\label{eq: effective diss S}
    \begin{aligned}
        S_d^\textup{eff}=&-\ii \frac{4\eta}{\gamma}\int \dd{t}\sum_{\substack{p,p^\prime \\ \sigma,\sigma^\prime}} \sigma\sigma^\prime V_p V_{p^\prime}\Big[b_-(t) \bar{c}_{p\sigma-}(t)\bar{f}_{\bar{\sigma}-}(t)\bar{b}_+(t) f_{\bar{\sigma}^\prime+}(t)c_{p^\prime\sigma^\prime+}(t)\\
        &-\frac{1}{2}b_+(t) \bar{c}_{p\sigma+}(t)\bar{f}_{\bar{\sigma}+}(t)\bar{b}_+(t-0) f_{\bar{\sigma}^\prime+}(t-0)c_{p^\prime\sigma^\prime+}(t-0)\\
        &-\frac{1}{2}b_-(t) \bar{c}_{p\sigma-}(t)\bar{f}_{\bar{\sigma}-}(t)\bar{b}_-(t+0) f_{\bar{\sigma}^\prime-}(t+0)c_{p^\prime\sigma^\prime-}(t+0)\Big]~,
    \end{aligned}
\end{equation}
where $t\pm0$ indicates an infinitesimal shift of the times that is necessary to correctly calculate tadpole contributions to the self-energies.
\par The non-crossing approximation (NCA) amounts to a self-consistent perturbation theory (i.e. a skeleton expansion of the self-energy) in the impurity-lead hopping $V_p$ truncated at the leading order \cite{Bickers,SlaveBosons,PRB_Nonequilibrium_NCA,transport_aim,PRB_noneq_aim,LangrethNordlander1,LangrethNordlander2}, namely the second. The residual dissipation needs to be included at the same order. Since the dissipative vertex \eq~\eqref{eq: effective diss S} is already of order $V_p^2$, we only need to include it at the mean-field (i.e. tadpole) level. The resulting approximation to the self-energy is then conserving \cite{PRB_Nonequilibrium_NCA,PRB_noneq_aim,LangrethNordlander1,LangrethNordlander2}. The Luttinger-Ward functional \cite{Fabrizio} truncated to the second order in the impurity-bath hopping is shown in \fig~\ref{fig: diagrams}, along with the relevant self-energies. 
\par The only nontrivial aspect of the slave boson mapping is that the constraint $Q=1$ has to be taken into account exactly. This can be done in a standard fashion \cite{Bickers,PRB_Nonequilibrium_NCA,transport_aim,PRB_noneq_aim,LangrethNordlander1,LangrethNordlander2} by endowing the auxiliary particles with a fictitious chemical potential that is taken to infinity to extract the physical observables. Following \rrefs~\cite{LangrethNordlander1,LangrethNordlander2}, we implement this constraint by discarding all terms that feature one or more lesser Green's function (bosonic or fermionic) in the retarded self-energy, and keeping at most one lesser Green's function in the expression of the lesser self-energy.   
\par We are going to derive the equations describing the time evolution of the system prepared in a factorized state $\rho_0=\chi_d\otimes\rho_l$ between the impurity $\chi_d$ and the leads $\rho_d$. For our purposes, the initial state of the impurity is completely characterized by the initial occupancy of the dot site $n_\sigma^0=\expval*{d_\sigma^\dag d_\sigma}_\chi$, which coincides with the auxiliary fermion occupancy $\expval*{f_\sigma^\dag f_\sigma}=\expval*{d_\sigma^\dag d_\sigma}$ (and determines the initial boson occupancy $\expval*{b^\dag b}_\chi=1-\sum_\sigma n_\sigma^0$ through the constraint). The leads are assumed to be prepared in their own equilibrium states $\rho_l=\bigotimes_\alpha\exp(-\beta H_\alpha)/Z_\alpha$, where $H_\alpha=\sum_{p\sigma\alpha}(\ce_p-\mu_\alpha)c_{p\sigma\alpha}^\dag c_{p\sigma\alpha}$ and $Z_\alpha=\Tr\exp(-\beta H_\alpha)$. The subsequent evolution can be thought as a quench in which the dot sites and the leads are initially disconnected (i.e. $V_p=0$) and suddenly put into contact at time $t=0$ by turning on the dot-leads hopping $V_p$. Hence, we will consider a time-dependent $V_p(t)=\theta(t) V_p$, which is necessary to set the correct limits in some integrals over time. The time dependence of the tunneling terms, i.e. the perturbation, makes all self energies $\Sigma_{f\sigma}(t,t^\prime)$, $\Pi(t,t^\prime)$ vanish whenever either of their time arguments is negative. Thus, we will always consider $t\geq0$, $t^\prime\geq0$ in the following equations, and we will consequently omit the $\theta$ functions in  front of the self-energies. 
\par The contour-ordered NCA self-energies associated to the unitary part of the dynamics of the auxiliary particles $f_\sigma$ and $b$ are given by \cite{Bickers,SlaveBosons,PRB_Nonequilibrium_NCA,transport_aim,PRB_noneq_aim,LangrethNordlander1,LangrethNordlander2} 
\begin{equation}
\begin{aligned}
    \Sigma_{f\sigma}(t,t^\prime)&=\ii\bm{\Delta}_\sigma(t,t^\prime) B(t,t^\prime)\quad\text{for auxiliary fermions,}\\
    \Pi (t,t^\prime)&=-\ii\sum_\sigma \bm{\Delta}_\sigma(t^\prime,t) G_\sigma(t,t^\prime)\quad\text{for bosons}
\end{aligned}
\end{equation}
where $G_\sigma(t,t^\prime)\equiv -\ii\expval{\mathcal{T}f_\sigma(t) f_\sigma^\dag(t^\prime)}$ and $B(t,t^\prime)\equiv -\ii\expval*{\mathcal{T}b(t) b^\dag(t^\prime)}$ are the contour-ordered Green's functions for the auxiliary fermions and the bosons, respectively ($\mathcal{T}$ is the contour-ordering symbol). The function $\bm{\Delta}_\sigma(t,t^\prime)$ is the local Green's function of the leads, namely $\bm{\Delta}_\sigma(t,t^\prime)\equiv \sum_{p,\alpha,p^\prime,\alpha^\prime}V_p(t) g_{p\sigma\alpha,p^\prime\sigma\alpha^\prime}(t,t^\prime)V_p(t^\prime)$, where $g_{p\sigma\alpha,p^\prime\sigma\alpha^\prime}(t,t^\prime)\equiv -\ii\expval*{\mathcal{T}c_{p\sigma\alpha}(t)c^\dag_{p^\prime\sigma\alpha^\prime}(t^\prime)}$. Applying Langreth's rules \cite{HaugJauho} and the projection onto $Q=1$ we obtain the various components of the self-energies:
\begin{equation}
    \begin{aligned}
        \Sigma_{f\sigma}^{\gtrless}(t,t^\prime)&=\ii\Delta_\sigma^{\gtrless}(t-t^\prime) B^{\gtrless}(t,t^\prime)~,\\
        \Sigma_{f\sigma}^{R,A}(t,t^\prime)&=\ii\Delta_\sigma^>(t-t^\prime) B^{R,A}(t,t^\prime)
    \end{aligned}
\end{equation}
and 
\begin{equation}
    \begin{aligned}
        \Pi^{\gtrless} (t,t^\prime)&=-\ii\sum_\sigma \Delta_\sigma^\lessgtr(t^\prime-t) G_\sigma^\gtrless(t,t^\prime)~,\\
        \Pi^{R,A} (t,t^\prime)&=-\ii\sum_\sigma \Delta_\sigma^<(t^\prime-t) G_\sigma^{R,A}(t,t^\prime)~.
    \end{aligned}
\end{equation}
The projection on the physical subspace has two effects on the self-energies. The first is that some contributions are discarded, while the second is that the local Green's function $\bm{\Delta}_\sigma(t,t^\prime)$ has to be substituted by its \emph{unperturbed} version $\Delta_\sigma(t,t^\prime)$. This occurs because any correction to $\Delta_\sigma(t,t^\prime)$ must contain at least one lesser Green's function of the fermions or the bosons \cite{LangrethNordlander1,LangrethNordlander2}, contrary to the projection rule. This substitution is not an approximation, and it does not imply that the leads are not affected by the impurity. In fact, we will show in the section devoted to transport that the knowledge of the Green's functions of the auxiliary particles is sufficient to compute $g_{p\sigma\alpha,p^\prime\sigma\alpha^\prime}(t,t^\prime)$.
\par Since the unperturbed leads are assumed to be in thermodynamic equilibrium, we have [remembering $V_p(t)=\theta(t)V_p$] $\Delta_\sigma(t,t^\prime)\equiv\theta(t)\theta(t^\prime)\Delta_\sigma(t-t^\prime)$ and we can express $\Delta_\sigma(t-t^\prime)$ through its Fourier transform $\Delta_\sigma(\omega)=\int\dd{t}\ee^{\ii \omega^+(t-t^\prime)}\Delta_\sigma(t-t^\prime)$ (where $\omega^\pm\equiv \omega\pm \ii 0$ denotes an infinitesimal shift along the imaginary axis):
$\Delta_\sigma(\omega)=\sum_\alpha \Delta_{\sigma\alpha}(\omega)$ where
\begin{subequations}
\begin{align}
    &\Delta^{R,A}_{\sigma\alpha}(\omega)=\int \frac{\dd{\ce}}{2\pi}\frac{\Gamma_\alpha(\ce)}{\omega^\pm-\ce}=\mathcal{P}\int \frac{\dd{\ce}}{2\pi}\frac{\Gamma_\alpha(\ce)}{\omega-\ce}\mp\frac{\ii}{2}\Gamma_\alpha(\omega)\\
    &\Delta^<_{\sigma\alpha}(\omega)=\ii\Gamma_\alpha(\omega)F_\alpha(\omega)\\
    &\Delta^>_{\sigma\alpha}(\omega)=-\ii\Gamma_\alpha(\omega)[1-F_\alpha(\omega)]
\end{align}
\end{subequations}
We have introduced the level width function $\Gamma_\alpha(\omega)\equiv2\pi\sum_p V_{p\alpha}^2\delta(\omega-\ce_p)$ and the Fermi distribution $F_\alpha(\omega)\equiv \{\exp[\beta(\omega-\mu_\alpha)]+1\}^{-1}$ of the lead $\alpha$, with inverse temperature $\beta$ \footnote{Since we are not interested in thermal transport, we assume that both baths have the same temperature.}. Notice that we keep the label $\sigma$ for the sake of generality, although we never consider the presence of magnetic fields or initially spin-unbalanced leads that could lead to an explicit spin dependence of $\Delta$. Also, we temporarily keep the label $\alpha$ on $\Gamma_\alpha$, although we will always consider two identical baths with $\Gamma_L=\Gamma_R$. 
\par In our calculations, we assume zero temperature $\beta\to\infty$ and we take a flat density of states $\Gamma(\omega)=\Gamma\xi(\omega)$, with $\xi(\omega)=\theta(W-\abs{\omega})$, for which
\begin{equation}
    \begin{aligned}
        \Delta^<_{\sigma\alpha}(t)&=-\frac{\Gamma}{2\pi}\frac{\ee^{-\ii\mu_\alpha t}-\ee^{\ii W t}}{t}\\
         \Delta^>_{\sigma\alpha}(t)&=\frac{\Gamma}{2\pi}\frac{\ee^{-\ii W t}-\ee^{-\ii\mu_\alpha t}}{t}
    \end{aligned}~.
\end{equation}
\par The dissipative contributions to the retarded self-energies are 
\begin{equation}
    \begin{aligned}
        \Sigma_{f\sigma}^{\textup{diss},R}(t,t^\prime)&=-\ii\frac{2}{\gamma}B^>(t,t)\bm{\Delta}^<_{\sigma}(t,t)\delta(t-t^\prime)\\
        &\overset{\textup{proj.}}{\to}-\ii\frac{2}{\gamma}[B^>(t,t)-B^<(t,t)]\Delta^<_{\sigma}(t,t)\delta(t-t^\prime)=-\frac{2}{\gamma}\Delta^<_{\sigma}(t,t)\delta(t-t^\prime)\equiv-\frac{\ii}{2}\kappa_\sigma\delta(t-t^\prime)\\
        \Pi^{\textup{diss},R}(t,t^\prime)&= -\frac{2\ii}{\gamma}\sum_\sigma\bm{\Delta}^<_\sigma(t,t)G_{\bar{\sigma}}^<(t,t)\delta(t-t^\prime)\overset{\textup{proj.}}{\to}0
    \end{aligned}
\end{equation}
where $\overset{\textup{proj.}}{\to}$ stands for the projection procedure of discarding all lesser fermionic or bosonic Green's functions, mentioned in the previous paragraphs. In the above Equations we introduced the effective dissipation rate
\begin{equation}
    \kappa_\sigma\equiv-\ii\frac{4}{\gamma}\Delta^<_{\bar{\sigma}}(t,t)=\sum_\alpha\frac{2\Gamma_{\alpha}}{\pi\gamma}\int_{-\infty}^{\infty}\dd{\omega}\xi(\omega)F_{\alpha}(\omega)~,
\end{equation}
where $\xi(\omega)$ is the cutoff function defined by $\Gamma_\alpha(\omega)=\Gamma_\alpha\xi(\omega)$. For a flat level width function $\xi(\omega)=\theta(W-\abs{\omega})$ we have 
\begin{equation}
\kappa_\sigma=2\sum_\alpha\frac{\Gamma_\alpha}{\pi\gamma}(\mu_\alpha+W)~,
\end{equation}
while for a general shape $\xi(\omega)$ we can estimate $\kappa\sim \Gamma/\gamma\cdot W$. Thus, the effective loss rate depends on the full band-shape of the leads, and grows with the bandwidth. 
\par The lesser self-energies read
\begin{equation}
    \begin{aligned}
        \Sigma_{f\sigma}^{\textup{diss},<}(t,t^\prime)&=0\\
        \Pi^{\textup{diss},<}(t,t^\prime)&=\frac{4\ii}{\gamma}\bm{\Delta}^<_\sigma(t,t)G_{\bar{\sigma}}^<(t,t)\delta(t-t^\prime)\overset{\textup{proj.}}{\to}\frac{4\ii}{\gamma}\sum_\sigma\Delta^<_\sigma(t,t)G_{\bar{\sigma}}^<(t,t)\delta(t-t^\prime)=-\sum_\sigma\kappa_\sigma G_{\sigma}^<(t,t)\delta(t-t^\prime)
    \end{aligned}
\end{equation}
As anticipated, the leading order self-energies from the effective interaction are mean-field-like corrections, because they are local in time.
\paragraph{Dyson equations} 
Using the self-energies written above, we can write down the Dyson equations determining the dynamics of the retarded Green's functions
\begin{equation}\label{eq: dyson G}
    \begin{aligned}
        (\ii\partial_t-\ce_\sigma)G_\sigma^R(t,t^\prime)&=\delta(t-t^\prime)-\frac{\ii}{2}\kappa_\sigma G^R_\sigma(t,t^\prime)+\ii\int_{t^\prime}^t\dd{\bar{t}}\Delta^>_\sigma(t-\bar{t})B^R(t,\bar{t})G^R_\sigma(\bar{t},t^\prime)~,\\
        (-\ii\partial_{t^\prime}-\ce_\sigma)G_\sigma^R(t,t^\prime)&=\delta(t-t^\prime)-\frac{\ii}{2}\kappa_\sigma G^R_\sigma(t,t^\prime)+\ii\int_{t^\prime}^t\dd{\bar{t}}G^R_\sigma(t,\bar{t})\Delta^>_\sigma(\bar{t}-t^\prime) B^R(\bar{t},t^\prime)
    \end{aligned}
\end{equation}
for the auxiliary fermions, and
\begin{equation}\label{eq: dyson B}
    \begin{aligned}
        \ii\partial_t B^R(t,t^\prime)&=\delta(t-t^\prime)-\ii\sum_\sigma\int_{t^\prime}^t\dd{\bar{t}}\Delta^<_\sigma(\bar{t}-t)G^R_\sigma (t,\bar{t})B^R(\bar{t},t^\prime)~,\\
        -\ii\partial_{t^\prime} B^R(t,t^\prime)&=\delta(t-t^\prime)-\ii\sum_\sigma\int_{t^\prime}^t\dd{\bar{t}}B^R(t,\bar{t})\Delta^<_\sigma(t^\prime-\bar{t})G^R_\sigma(\bar{t},t^\prime)
    \end{aligned}
\end{equation}
for the bosons. The advanced Green's functions can be obtained as $G^A(t,t^\prime)=[G^R(t^\prime,t)]^\ast$, with $G=G_\sigma$ or $B$. The boundary conditions for the Dyson equations are $G^R_\sigma(t+0,t)=B^R(t+0,t)=-\ii$, which simply enforces the (anti-)commutation relations. A peculiar behavior of the Dyson equations in the NCA is that they only involve retarded functions, which greatly simplifies their numerical solution. In particular, the retarded Green's function that are obtained are time-translational invariant: $G_\sigma^R(t,t^\prime)\equiv G_\sigma^R(t-t^\prime)$, $B^R(t,t^\prime)\equiv B^R(t-t^\prime)$. For $\kappa=0$ we obtain the same solution as in equilibrium. If this behavior might seem unphysical, let us recall that it regards the dynamics of two unphysical, auxiliary particles $f_\sigma$ and $b$ of which the physical fermion $d_\sigma=b^\dag f_\sigma$ [cf. \eqs~\eqref{eq: slave bosons}] is “composed”. Indeed, the object that we aim to compute is the physical Green's function for the $d_\sigma$ operators, which in the language of the slave boson representation is a four-point vertex. For this object, the time evolution of the retarded and lesser components are intertwined. 
\paragraph{Kinetic equations}
The lesser components of the auxiliary fermion Green's functions are determined by
\begin{equation}\label{eq: lesser G f}
    \begin{aligned}
        (\ii\partial_t-\ce_\sigma)G^<_\sigma(t,t^\prime)&=-\frac{\ii}{2}\kappa_\sigma G_\sigma^<(t,t^\prime)+\ii[(\Delta^>_\sigma B^R)\ast G^<_\sigma](t,t^\prime)+\ii[\Delta^<_\sigma B^<\ast G^A_\sigma](t,t^\prime)~,\\
        (-\ii\partial_{t^\prime}-\ce_\sigma)G^<_\sigma(t,t^\prime)&=+\frac{\ii}{2}\kappa_\sigma G_\sigma^<(t,t^\prime)+[G^R_\sigma\ast\ii\Delta^<_\sigma B^<](t,t^\prime)+[G^<_\sigma\ast\ii\Delta^>_\sigma B^A](t,t^\prime)~.
    \end{aligned}
\end{equation}
The above equations have been written in a compact notation that treats the Green's functions as matrices in the two time indices. We have introduced two different “products” for the Green's functions: a direct product $(AB)(t,t^\prime)=A(t,t^\prime) B(t,t^\prime)$ and the time convolution (i.e. the matrix product) $(A\ast B)(t,t^\prime)\equiv\int\dd{\bar{t}}A(t,\bar{t}) B(\bar{t},t^\prime)$. The time evolution of the diagonal component reads
\begin{equation}\label{eq: lesser G f diag}
\begin{aligned}
     \ii \mathrm{d}_t & G^<_\sigma(t,t)=[\ii\partial_t G^<_\sigma(t,t^\prime)+\ii\partial_{t^\prime} G^<_\sigma(t,t^\prime)]\big\rvert_{t^\prime=t}=\\
     &=-\ii\kappa_\sigma G^<_\sigma(t,t)+2\Re\Big\{\ii\int_0^t\dd{\bar{t}}\Big[\Delta^<_\sigma(t-\bar{t})B^<(t,\bar{t})G^A_\sigma(\bar{t},t)+\Delta^>_\sigma(t-\bar{t})B^R(t,\bar{t})G^<_\sigma(\bar{t},t)\Big]\Big\}
\end{aligned}
\end{equation}
The boundary conditions are simply the initial dot populations: $G_\sigma(0,0)=\ii n_{d\sigma}(0)$. \par For the bosons we similarly obtain
\begin{equation}\label{eq: lesser G b}
    \begin{aligned}
        \ii\partial_t B^<(t,t^\prime)=&-\sum_\sigma\kappa_\sigma G_\sigma^<(t,t) B^A(t-t^\prime)\\
        &-\ii\sum_\sigma[(\Delta^<_\sigma)^T G_\sigma^R\ast B^<](t,t^\prime)-\ii\sum_\sigma[(\Delta^>_\sigma)^T G_\sigma^<\ast B^A](t,t^\prime)~,\\
        -\ii\partial_{t^\prime}B^<(t,t^\prime)=&-\sum_\sigma\kappa_\sigma G_\sigma^<(t,t) B^R(t-t^\prime)\\
        &-[B^R\ast\ii\sum_\sigma(\Delta^>_\sigma)^T G_\sigma^<](t,t^\prime)-[B^<\ast\ii\sum_\sigma(\Delta^<_\sigma)^T G_\sigma^A](t,t^\prime)
    \end{aligned}
\end{equation}
where we employ the notation $(\Delta^T)(t,t^\prime)=\Delta(t^\prime,t)$. To find the equation determining the evolution of the diagonal component one uses $\lim_{t^\prime\to t}[B^A(t,t^\prime)-B^R(t,t^\prime)]=\lim_{t^\prime\to t}[B^<(t,t^\prime)-B^>(t,t^\prime)]=-\ii\comm{b^\dag}{b}=\ii$ and finds
\begin{equation}\label{eq: lesser G b diag}
\begin{aligned}
     \ii \mathrm{d}_t B^<(t,t)=&-\ii\sum_\sigma\kappa_\sigma G^<_\sigma(t,t)\\
     &+2\Re\Big\{\ii\sum_\sigma\int_0^t\dd{\bar{t}}\Big[\Delta^<_\sigma(t-\bar{t})B^<(t,\bar{t})G^A_\sigma(\bar{t},t)+\Delta^>_\sigma(t-\bar{t})B^R(t,\bar{t})G^<_\sigma(\bar{t},t)\Big]\Big\}~.
\end{aligned}
\end{equation}
It is easy to observe that the constraint is explicitly conserved: $\mathrm{d}_t Q(t)=\ii \mathrm{d}_t B^<(t,t)-\sum_\sigma \ii \mathrm{d}_t G^<_\sigma(t,t)=0$. So, if we take the appropriate initial condition $B(0,0)=-\ii [1-\sum_\sigma n_{d\sigma}(0)]$, we will always keep $Q=1$ during the dynamics, within the numerical accuracy of our computations. 
\begin{figure}
    \centering
    \includegraphics[width=.35\linewidth]{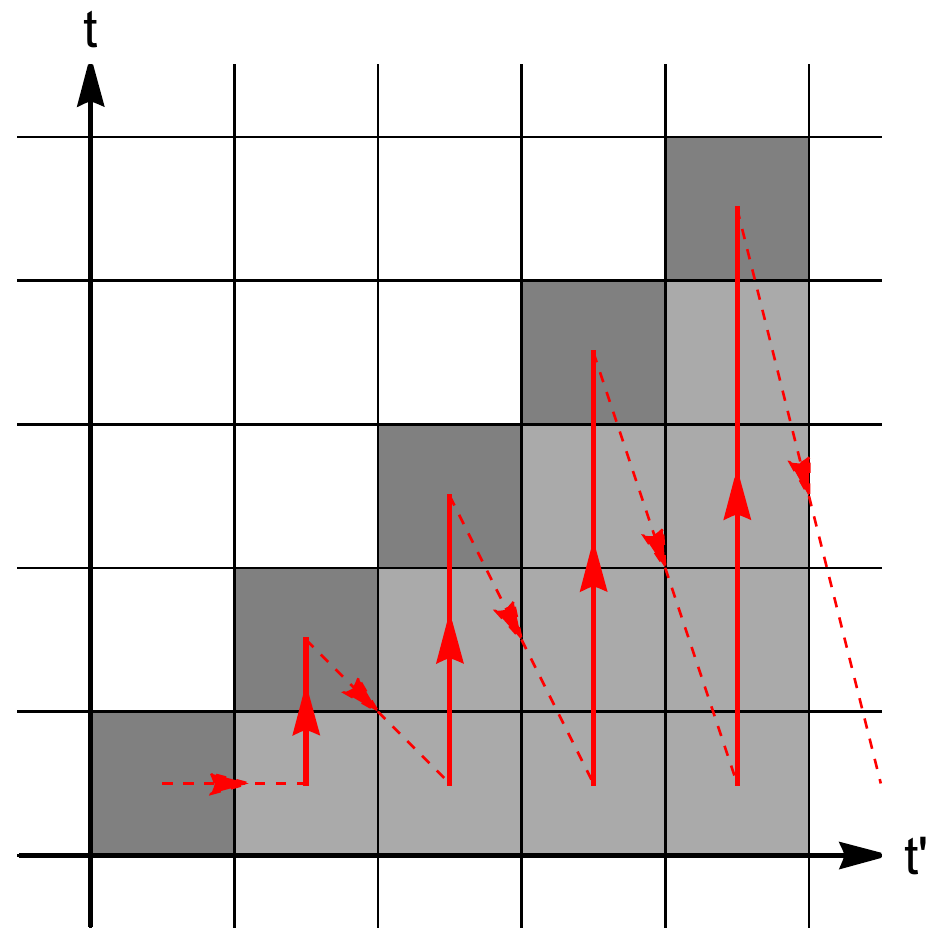}
    \caption{Sketch of the steps for the numerical integration for the determination of the lesser Green's functions. The first square on the diagonal is the initial condition $G_\sigma^<(0,0)=\ii n_\sigma^0$, $B^<(0,0)=1-\sum_\sigma n_\sigma^0$, and the subsequent integrations proceed along the columns, from $t^\prime=0$ to $t^\prime=t$ (thick red arrows). The steps corresponding to the light gray squares make use of \eqs~\eqref{eq: lesser G f} and \eqref{eq: lesser G b}, while the diagonal elements in darker gray employ \eqs~\eqref{eq: lesser G f diag} and \eqref{eq: lesser G b diag}.}
    \label{fig: integration contour}
\end{figure}
\paragraph{Numerical implementation} We solve the Equations above using the simple algorithm described in \rref~\cite{LangrethNordlander2}. Conceptually, we have equations in the form $\dv{}{t}f(t)=K[f](t)$, where the right-hand side depends functionally on $f$. We first convert the equation to an integral one to obtain one integration step: $\int_t^{t+\delta t}\dd{t^\prime}\dv{}{t^\prime}f(t^\prime)=f(t+\delta t)-f(t)=\int_t^{t+\delta t}\dd{t^\prime}K[f](t^\prime)$. The integrals on the right-hand side (i.e. the one for the time step and the one implied in $K$) are computed by any quadrature rule. In our case, the simple trapezoidal rule with equally spaced points suffices. This choice leads to clearer equations, without sacrificing too much the accuracy (at the cost of needing a $\delta t$ which cannot be too large). A key point is that the update rule for $f(t+\delta t)$ is implicit, since $f(t+\delta t)$ appears also in the discretization of the right-hand side. This makes the integration numerically stable. A nice feature of the implementation of \rref~\cite{LangrethNordlander2} is that $Q$ is exactly conserved also in the discretized dynamics, which implies that the deviations from $1$ will be comparable by machine precision $\sim 10^{-14\divisionsymbol 15}$.
\par The Dyson \eqs~\eqref{eq: dyson G} and \eqref{eq: dyson B} are integrated first, since they do not depend on the lesser functions. The equations for the fermionic and bosonic Green's functions are coupled, though, and need to be solved together. We only integrate the forward equations $\ii\partial_t G_\sigma(t-t^\prime)=\dots$, etc., and we take advantage of the time-translational invariance of the retarded Green's functions. The retarded functions thus obtained are then used to compute the lesser functions. The causal properties of the equations implies that we need to follow the path depicted in \fig~\ref{fig: integration contour} on the discretized time grid $(t,t^\prime)$. Since lesser Green's functions are anti-Hermitian matrices in the time indices \cite{Kamenev,HaugJauho}, we compute only their values for $t\leq t^\prime$. 
\par In most of our calculations we set the time step to $\delta t=0.5 W^{-1}$. We have verified that doubling or halving this value leads to differences of at most a few per mil in Green's functions and currents. The value of the observables after the initial transient, i.e. for times larger than a few $\Gamma^{-1}$, are less susceptible to changes of $\delta t$---a possible manifestation of the dynamically attractive nature of the local stationary state.
\subsection{Physical Green's function}
In the subspace with no double occupancies $d_\sigma=X_{0\sigma}=b^\dag f_\sigma$. Hence 
\begin{equation}
    G_{d\sigma}(t,t^\prime)=-\ii\expval{\mathcal{T}_\mathcal{C}d_\sigma(t)d_\sigma^\dag(t^\prime)}=-\ii\expval{\mathcal{T}_\mathcal{C}f_\sigma(t) b^\dag(t)b(t^\prime)f_\sigma^\dag(t^\prime)}~.
\end{equation}
Within the non-crossing approximation \cite{Bickers,slaveBosons_Barnes,SlaveBosons,PRB_noneq_aim,PRB_Nonequilibrium_NCA,transport_aim,LangrethNordlander1,LangrethNordlander2} one simply decouples the fermions from the bosons (i.e. one ignores vertices):
\begin{equation}
    G_{d\sigma}(t,t^\prime)\approx \ii G_\sigma(t,t^\prime)B(t^\prime,t)~.
\end{equation}
This approximation can be obtained as the leading term in a large-$N$ expansion in the number of flavors of fermions (with the rescaling $V_p\to V_p/N^{1/2}$). While the auxiliary particles' Green functions are approximated with a good accuracy by the NCA equations derived above, the neglect of vertices in the physical Green's function does introduce some spurious effects \cite{MullerHartmann,ComparisonNCA-NRG}. Nevertheless, it is known to provide the correct qualitative properties of the Kondo effect, with quantitative discrepancies of the order of $15\%$ for observables such as the zero-bias conductance \cite{transport_aim}. 
We obtain the lesser function as follows:
\begin{equation}
    G_{d\sigma}^<(t,t^\prime)\approx \ii G_\sigma^<(t,t^\prime)B^>(t^\prime,t)\overset{\textup{proj.}}{\to}\ii G_\sigma^<(t,t^\prime)[B^>(t^\prime,t)-B^<(t^\prime,t)]= G_\sigma^<(t,t^\prime)b(t^\prime,t)~,
\end{equation}
where the first approximate equality is the NCA, while the second equality is the projection onto the physical subspace $Q=1$. We have introduced the functions
\begin{equation}
    \begin{aligned}
        g_\sigma(t,t^\prime)&\equiv \ii[G_\sigma^>(t,t^\prime)-G_\sigma^<(t,t^\prime)]\\
        b(t,t^\prime)&\equiv \ii[B^>(t,t^\prime)-B^<(t,t^\prime)]\\
    \end{aligned}
\end{equation}
so that 
\begin{equation}
    \begin{aligned}
        G^R_\sigma(t,t^\prime)&=-\ii\theta(t-t^\prime)g_\sigma(t-t^\prime)\\
        G^A_\sigma(t,t^\prime)&=+\ii\theta(t^\prime-t)g_\sigma(t,t^\prime)
    \end{aligned}
\end{equation}
and analogously for the bosons. In particular, $b(t,t)=1$, so $\expval{d_\sigma(t)^\dag d_\sigma(t)}\equiv -\ii G_{d\sigma}^<(t,t)=-\ii G_{\sigma}^<(t,t)=\expval{f_\sigma(t)^\dag f_\sigma(t)}$, where the equality $G^<_{d\sigma}(t,t)=G^<_\sigma(t,t)$ is exact. Similarly,
\begin{equation}
    G_{d\sigma}^>(t,t^\prime)\approx \ii G_\sigma^>(t,t^\prime)B^<(t^\prime,t)\overset{\textup{proj.}}{\to}\ii [G_\sigma^>(t,t^\prime)-G_\sigma^<(t,t^\prime)]B^<(t^\prime,t)= g_\sigma(t,t^\prime)B^<(t^\prime,t)~,
\end{equation}
and finally
\begin{equation}\label{eq: physical GF}
    G_{d\sigma}^R(t,t^\prime)\equiv\theta(t-t^\prime)\big[G_{d\sigma}^>(t,t^\prime)-G_{d\sigma}^<(t,t^\prime)\big]=\ii\big[G^R_\sigma(t,t^\prime)B^<(t^\prime,t)+G_\sigma^<(t,t^\prime)B^A(t^\prime,t)\big]
\end{equation}
At first sight, there appears to be a problem with the anticommutation relations $\{d_\sigma,d^\dag_\sigma\}=\one$: $ G_{d\sigma}^>(t,t)=B^<(t,t)=-\ii+\sum_\sigma G_{\sigma}^<(t,t)\neq -\ii+G^<_{\sigma}(t,t)$, therefore
\begin{equation}
    G_{d\sigma}^R(t+0,t)=B^<(t,t)-G^<_\sigma(t,t)=-\ii(1-G^<_{-\sigma}(t,t))\neq -\ii\{d_\sigma,d^\dag_\sigma\}=-\ii
\end{equation}
This is not a shortcoming of the NCA, but rather of the Hilbert space truncation $d_\sigma\to X_{0\sigma}$. Indeed, $-\ii\{X_{0\sigma},X_{\sigma0}\}=-\ii(X_{00}+X_{\sigma\sigma})\neq\one$. This modified anticommutation relation means that the spectral function will not integrate to $1$, but rather to $\ii(B^<(t,t)-G^<_\sigma(t,t))=1-n_{-\sigma}(t)$. The missing spectral weight is that of double occupancies, which would appear as a very broad and low peak at $\omega=\ce_d$, with width $\sim\gamma$ and height $\sim \gamma^{-1}$.
\par In the main text, we show the dot spectral function in the local stationary state. This function is defined as $A_\sigma(\omega,t)=-\Im G_{d\sigma}^R(\omega,t)/\pi$, where $G_{d\sigma}^R(\omega,t)\equiv \int_0^\infty \dd{\tau}\ee^{\ii\omega\tau} G^R_{d\sigma}(t+\tau,t)$, choosing $t\gtrsim 5\Gamma_T^{-1}$ such that the $A_\sigma(\omega,t)$ has already saturated to its late-time value.
\subsection{Transport}
As usual in impurity problems \cite{HaugJauho}, the Green's function of the leads $g_{p\sigma\alpha,p^\prime\sigma\alpha^\prime}(t,t^\prime)$ is determined by a T-matrix equation 
\begin{equation}
    \hat{g}= \hat{g}_0+\hat{g}_0\ast \hat{P}\ast \hat{g}_0~,
\end{equation}
where the hat indicates a matrix in $p,\,\sigma,\,\alpha$ and times, and with the convolution symbol $\ast$ representing a matrix multiplication in all of the above indices.
The T-matrix $\hat{P}$ on the contour can be obtained by deriving the Luttinger-Ward functional (\fig~\ref{fig: diagrams}) and reads
\begin{equation}
    P_{p\sigma\alpha,p^\prime\sigma^\prime\alpha^\prime}(t,t^\prime)= V_{p\alpha}V_{p^\prime\alpha^\prime}\delta_{\sigma\sigma^\prime}\Big[\ii G_\sigma(t,t^\prime)B(t^\prime,t)+\frac{4}{\gamma} G_\sigma^<(t,t)\delta(t-t^\prime)\Big]~.
\end{equation}
When applying Langreth's rules to determine the various components of $\hat{P}$ we must apply the projection \cite{LangrethNordlander1,LangrethNordlander2} on the physical subspace $Q=1$. In this case there is a subtlety: the rule of keeping at most one auxiliary bosonic or fermionic lesser function applies only if the projected quantity vanishes in the unphysical $Q=0$ space. This is not the case for $\hat{g}$, since $\hat{g}_{Q=0}=\hat{g}_0\neq0$. Therefore, the projection must be done on $\hat{g}-\hat{g}_0$, and we obtain that $P=\order{G_\sigma^<,B^<}$ for all components. We can write the results in terms of a new function $Q_\sigma(t,t^\prime)$ (not to be confused with the charge $Q$) defined by
\begin{equation}
     P_{p\sigma\alpha,p^\prime\sigma^\prime\alpha^\prime}(t,t^\prime)= V_{p\alpha}V_{p^\prime\alpha^\prime}\delta_{\sigma\sigma^\prime} Q_\sigma(t,t^\prime)~.
\end{equation}
Then, within the NCA we have $Q_\sigma^{R,<}(t,t^\prime)=G_{d\sigma}^{R,<}(t,t^\prime)+Q_\sigma^{\textup{diss},R,<}(t)\delta(t-t^\prime)$, with 
\begin{equation}
    \begin{aligned}
         Q_\sigma^{\textup{diss},R}(t)&=-\frac{2}{\gamma} G_{\bar{\sigma}}^<(t,t)=-\frac{2\ii}{\gamma}n_{\bar{\sigma}}(t)\\
         Q_\sigma^{\textup{diss},<}(t)&=0
    \end{aligned}
\end{equation}
The current leaving reservoir $\alpha$ is defined as 
\begin{equation}
I_\alpha\equiv-\dv{}{t}\sum_{p\sigma}\expval*{c_{p\sigma\alpha}^\dag c_{p\sigma\alpha}}_t=-\dv{}{t}\sum_{p\sigma}g^<_{p\sigma\alpha,p\sigma\alpha}(t,t)~.    
\end{equation}
With the help of the appropriate T-matrix equation $\hat{g}^<= \hat{g}_0^<+\hat{g}_0^<\ast \hat{P}^A\ast \hat{g}_0^A+\hat{g}_0^R\ast \hat{P}^<\ast \hat{g}_0^A+\hat{g}_0^R\ast \hat{P}^R\ast \hat{g}_0^<$ and the NCA expression for $\hat{P}$ and $\hat{Q}$ we find
\begin{equation}\label{eq: currents}
\begin{aligned}
    I_\alpha(t)&=2\Re\int_0^t \dd{\bar{t}}\sum_\sigma\Big[Q_\sigma^R(t,\bar{t})\Delta_{\sigma\alpha}^<(\bar{t}-t)+Q_\sigma^<(t,\bar{t})\Delta_{\sigma\alpha}^A(\bar{t}-t)\Big]=\\
    &=2\Re\int_0^t \dd{\bar{t}}\sum_\sigma\Big[G^R_{d\sigma}(t,\bar{t})\Delta_{\sigma\alpha}^<(\bar{t}-t)+G_{d\sigma}^<(t,\bar{t})\Delta_{\sigma\alpha}^A(\bar{t}-t)\Big]+\frac{4}{\gamma}\sum_\sigma n_{\bar{\sigma}}(t)(-\mathrm{i})\Delta^<_{\sigma\alpha}(0)~.
\end{aligned}
\end{equation}
The first term in the second line of the above Equation is the usual starting point of the Meir-Wingreen formula \cite{MeirWingreenFormula,HaugJauho}, whereas the second term represents a new, dissipative contribution. We will show later that this latter contribution is related to the loss of particles from the system---a conclusion that can be anticipated by its proportionality to the dot occupancy. The loss term is proportional to the Green's function of the leads evaluated at coinciding times, $-\ii\Delta^<_{\sigma\alpha}(0)$, and is therefore non-universal, in the sense that it is sensitive to the full band shape of the leads, $\xi(\omega)$. For an infinite bandwidth $W$, this term would diverge \footnote{This limit cannot be taken, however, since the derivation of the effective Hamiltonian assumes $\gamma\gg W$.}.
\par We can then easily derive the conductance at zero bias, defined as $g(0,t)\equiv\lim_{\Delta\mu\to0}\dv*{I(t)}{\Delta\mu}$, where the transport current is $I(t)\equiv (I_L(t)-I_R(t))/2$, assuming that the left reservoir is at a higher chemical potential, $\mu_{L,R}=\mu\pm\Delta\mu/2$. Using 
\begin{equation}
    \dv{}{\Delta\mu}F_{L,R}(\omega)=\dv{}{\Delta\mu}F\Big[\omega-\big(\mu\pm\tfrac{\Delta\mu}{2}\big)\Big]=\mp\frac{1}{2} F^\prime(\omega-\mu_\alpha)
\end{equation}
in the expression for $\Delta_{\sigma\alpha}^<$ (with the prime $F^\prime$ indicating the derivative), we obtain at zero temperature
\begin{equation}\label{eq: conductance}
    g(0,t)=-\frac{\Gamma(\mu)}{2}\sum_\sigma\frac{1}{\pi}\Im\int_0^t\dd{\bar{t}}\ee^{\ii\mu\bar{t}}G_{d\sigma}^R(t,t-\bar{t})+\frac{\Gamma(\mu)}{\pi \gamma}n_d(t)~,
\end{equation}
which is \eq~(2) quoted in the main text. As noted there, the first term directly probes the spectral function at the chemical potential, and at later times converges to the usual stationary formula \cite{MeirWingreenFormula,HaugJauho} $g_\infty(0)=2^{-1}\Gamma(\mu) \sum_\sigma A_\sigma(\mu)$. Thus, the zero-bias conductance allows to detect directly the presence of the Kondo peak. Moreover, it is interesting to notice that the first term of \eq~\eqref{eq: conductance} only probes the spectral function at the chemical potential of the leads, thus displaying universal behavior in the usual Hamiltonian case. On the other hand, the correction represented by the second term explicitly depends on the dot population $n_d(t)$, which is non-universal (i.e. explicitly dependent on the cutoff $W$ and band shape $\xi(\omega)$). On the other hand, it does not diverge for an infinite bandwidth, unlike the individual currents $I_\alpha(t)$. Indeed, the “loss” terms of the currents, i.e. the last term of \eqref{eq: currents}, partially cancel out in the transported current even for a finite bias:
\begin{equation}
    \frac{2}{\gamma}\sum_\sigma n_{\bar{\sigma}}(t)(-\mathrm{i})[\Delta^<_{\sigma L}(0)-\Delta^<_{\sigma R}(0)]=\frac{2}{\gamma}n_d(t)\int\frac{\dd{\omega}}{2\pi}\Gamma(\omega)\big[F_L(\omega)-F_R(\omega)]=\frac{2}{\gamma}n_d(t)\int_{\mu_R}^{\mu_L}\frac{\dd{\omega}}{2\pi}\Gamma(\omega)~,
\end{equation}
where the last equality applies to zero temperature. In words, the losses contribute to the transport current only through the lead fermions present between the two chemical potentials. We need to remark that the loss correction to the zero-bias conductance is tiny with respect to the transport term (i.e. the first one of \eq~\eqref{eq: conductance}): while the latter is of order $10^{-1\divisionsymbol 0}$, the former is at most (for the maximal $n_d(t)=1$) $\Gamma/(\pi\gamma)$, which in our calculation is always less than $3\cdot 10^{-3}$. Thus, almost all of the effect of the residual dissipation affects the conductance through the impurity spectral function, and in particular the decrease of the height of the Kondo peak for decreasing dissipation rate.
\par It is interesting to compare the behavior of the transported current with the current of particles lost from the system, $I_\text{loss}(t)=-\dv*{[N_R(t)+N_L(t)+n_d(t)]}{t}=I_R(t)+I_L(t)-\dv*{n_d(t)}{t}$. Using the expressions \eqref{eq: currents}, \eqref{eq: lesser G f diag} and \eqref{eq: physical GF} we find
\begin{equation}\label{eq: I loss}
I_\textup{loss}(t)=\frac{4}{\gamma}\sum_\sigma n_{\bar{\sigma}}(t)(-\mathrm{i})\Delta^<_{\sigma}(0)=2\sum_\sigma \kappa_\sigma n_\sigma(t)~.
\end{equation}
As a check, we notice that in the limit $\gamma\to+\infty$ $\kappa_\sigma$ vanishes and we recover the statement of conservation of the number of particles: $I_\textup{loss}=0$. In general, after a transient of a few $\Gamma^{-1}$ the dot population saturates, and for a finite $\gamma$ the loss current saturates to $I_\textup{loss}(t)\to I_\textup{loss}^\infty=2\sum_\sigma \kappa_\sigma n_\sigma^\infty$. As it could be expected on classical grounds, the stationary loss current is proportional to the dot population and to the density of leads fermions at the dot site, since $\kappa_\sigma\propto\Delta^<_\sigma(0)$. The factor of $2$ accounts for the fact that each loss event entails the disappearance of two particles, one from the dot and one from the leads. In the large-$\gamma$ regime that we are analyzing, the losses are suppressed by an explicit factor of $\gamma$ (coming from $\kappa_\sigma$), besides of the slow increase of $n_d^\infty$ with $\gamma$ (see later). As we have already remarked, the dot population does not show universal behavior, hence the loss current does not provide a good probe of the Kondo effect emerging at large dissipation.
\par We can use $I_\textup{loss}(t)$ to estimate the double occupancies $\expval*{d_\uparrow^\dag d_\uparrow d_\downarrow^\dag d_\downarrow}$ that are still present in the Kondo regime. We compute $I_\textup{loss}(t)$ in the full theory:
\begin{equation*}
    I_\textup{loss}(t)=-\dv{}{t}\expval{N_\textup{tot}}=-\frac{\gamma}{2}\expval{\comm{L^\dag}{N_\textup{tot}}L+L^\dag\comm{L}{N_\textup{tot}}}
\end{equation*}
where we took into account that the Hamiltonian part of the dynamics conserves the total number of fermions $N_\textup{tot}=n_d+\sum_\alpha N_\alpha$. Using $\comm{N_\textup{tot}}{L}=\comm{n_d}{L}=-2L$, we obtain:
\begin{equation}
    I_\textup{loss}(t)=2\gamma\expval{L^\dag L}(t)=2\gamma\expval*{d_\uparrow^\dag d_\uparrow d_\downarrow^\dag d_\downarrow}(t)~.
\end{equation}
Hence, the number of double occupancies is
\begin{equation}
    \expval*{d_\uparrow^\dag d_\uparrow d_\downarrow^\dag d_\downarrow}(t)=\frac{I_\textup{loss}(t)}{2\gamma}=\frac{1}{\gamma}\sum_\sigma \kappa_\sigma n_\sigma(t)~,
\end{equation}
where in the last equality we used \eq~\eqref{eq: I loss} from the effective model at large $\gamma$---we are assuming that the results of the full theory should converge smoothly to those of the effective one in the regime of validity of the latter. Recalling that $\kappa_\sigma\sim \order{\Gamma_T W/\gamma}$ and that $n_d(t)$ has only a weak dependence on $\gamma$, we estimate that in the Kondo regime $\expval*{d_\uparrow^\dag d_\uparrow d_\downarrow^\dag d_\downarrow}(t)\sim \Gamma_T W/\gamma^2$ is suppressed as $\gamma^{-2}$.
\section{Additional data}
\begin{figure}
    \centering
    \includegraphics[width=.5\linewidth]{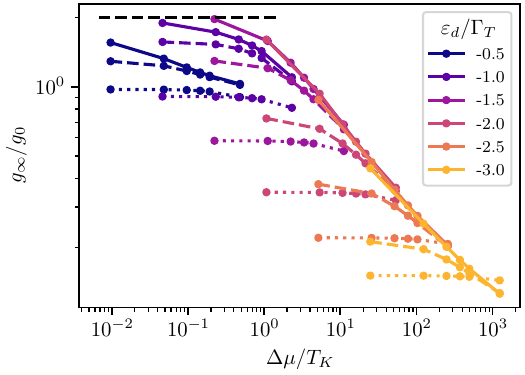}
    \caption{Loss of scaling collapse in the nonlinear conductance in presence of a finite dissipation. The data are for $\Gamma_T=0.02 W$, $T_K=(\Gamma_T W/2)^{1/2}\exp(\pi\ce_d/\Gamma_T)$. The full curves are for the AIM at $\gamma\to+\infty$, the dashed curves are for $\gamma=10W$ and the dotted curves are for $\gamma=W$. The black dashed line marks the maximal conductance value $2 g_0$. Decreasing $\gamma$, besides decreasing the conductance itself, spoils the data collapse as a function of $\Delta\mu/T_K$.}
    \label{fig: conductance}
\end{figure}

In this Section, we provide additional data on the relation between the loss rate $\gamma$ and the presence of typical signatures of the Kondo effect.
\par In \fig~\ref{fig: conductance} we show the long-time nonlinear conductance $g_\infty=\lim_{t\to\infty}I(t)/\Delta\mu$ as a function of the ratio $\Delta\mu/T_K$. A clear signature of the Kondo effect for $\gamma\to+\infty$ (continuous lines) is that the curves from different dot energies $\ce_d$ collapse on a universal curve (except those in the mixed-valence regime $\ce_d\gtrsim-\Gamma_T$) \cite{Hewson,Bickers,transport_aim,transport_aim,PRB_noneq_aim}. This scaling collapse signals the presence of $T_K$ as the only energy scale governing the low-energy transport properties, and can be used to detect the Kondo effect. In the Figure, dashed and dotted lines refer to the finite-dissipation cases of $\gamma=10W$ and $\gamma=W$, respectively. We observe that in this regime the scaling collapse is lost, since the new competing energy scale $\kappa_\sigma$ emerges. Especially for the smaller $\gamma=W$, the conductance becomes scarcely dependent on the bias $\Delta\mu$. From the central panel of \fig~(1) in the main text, this mild dependence can be attributed to the complete loss of the Kondo peak in the spectral function, meaning that the dot behaves simply as a noninteracting dot, which has its maximal sensitivity to the bias only for $\mu\approx\ce_d$. 
\begin{figure}
    \centering
    \includegraphics[width=\linewidth]{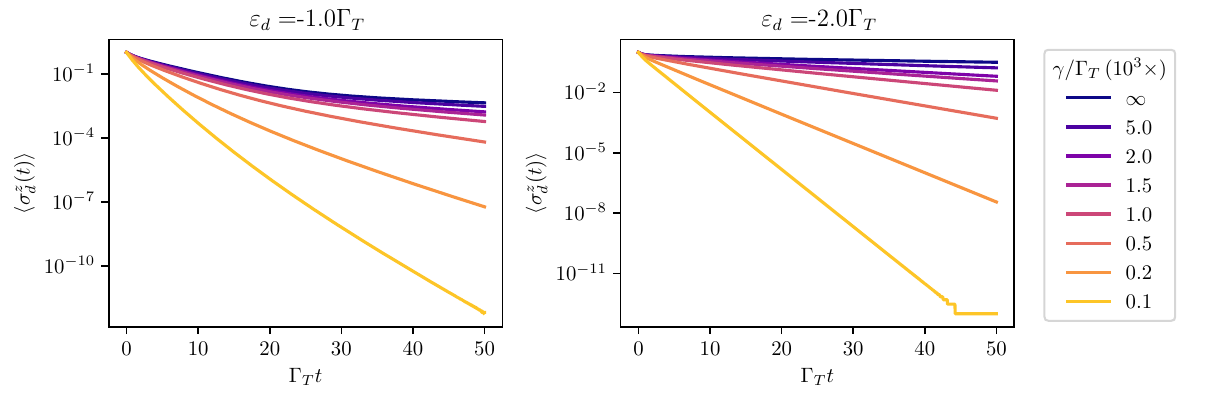}
    \caption{Spin decay in real time for a dot in the mixed-valence regime (left) and in the Kondo regime (right). In both scenarios we can observe how the decay is suppressed by an increasing loss rate $\gamma$, thus signaling the regime of strong correlations. Both plots are for $\Gamma_T=10^{-2} W$. The discrete steps in the $\gamma=100\Gamma_T$ curve of the right plot are caused by the reaching of machine precision $\expval*{\sigma_d^z(t)}\sim 10^{-13}.$}
    \label{fig: spin decay}
\end{figure}
\par In \fig~\ref{fig: spin decay} we show the typical appearance of the spin decay in the mixed-valence regime $\ce_d=-\Gamma_T$ (left) and in the Kondo regime $\ce_d=-2\Gamma_T$ (right). While for $\abs{\ce_d}\lesssim\Gamma_T$ the decay is slower than exponential, for deeper dot levels a purely exponential form is recovered. Regardless of the form of the decay, the effect of the two-body loss is to suppress the equilibration of the impurity spin, down to quite a small value. This suppression signals the crossover to the strongly dissipated regime in which the physics is dominated by the Kondo resonance (which corresponds precisely to a long-lived spin state \cite{Coleman}). The curves in the rightmost panel of \fig~(1) in the main text are obtained via a linear fit of $\log\expval*{\sigma_d^z(t)}$ versus time, excluding the initial data for $t\le 10\Gamma_T^{-1}$. According to the literature \cite{PRB_spin_decay1,PRL_spin_decay1,MPS_AIM_Wauters}, the exponential decay rate of $\expval*{\sigma_d^z(t)}$ in the Kondo regime should be proportional to the Kondo temperature---namely, it should scale exponentially with the ratio $\ce_d/\Gamma_T$ with a slope of $\pi$. While in the right panel of \fig~(1) in the main text we do observe an exponential scaling, we find that the slope is smaller, being close to $0.56\pi$. A similar behavior is observed with the variational method in the companion paper \cite{companion}.
\section{Discussion of the many-sites setup}
In this Section, we provide a detailed discussion of the setup in which the dot is composed of more than one site, and on the conditions for the mapping of this dissipative model to a higher-spin Kondo model.
\par We consider the Lindblad master equation for a composite dot made of $\ell_d$ sites, each one subject to a two-body decay rate of $\gamma_j$:
\begin{equation}
    \dv{}{t}\rho(t)=-\ii\comm{H}{\rho(t)}+\sum_{j\in\textup{dot}}\gamma_j\Big(L_j\rho(t)L_j^\dag-\frac{1}{2}\{L_j^\dag L_j,\rho(t)\}\Big)~,
\end{equation}
with jump operators $L_j\equiv d_{j\downarrow} d_{j\uparrow}$ (the operator $d_{j\sigma}$ annihilates a fermion with spin $\sigma$ at the dissipative site $j$) and a Hamiltonian having the usual form $H=H_\textup{dot}+H_\textup{leads}+H_\textup{tun}$, with
\begin{equation}
    \begin{aligned}
    H_\textup{dot}&=\sum_{i,j\in\textup{dot},\sigma}h_{ij}d_{i\sigma}^\dag d_{j\sigma}\\
        H_\textup{leads}&=\sum_{p\sigma\alpha} \ce_{p\alpha} c_{p\sigma\alpha}^\dag c_{p\sigma\alpha}\\
    H_\textup{tun}&=\sum_{pj\alpha\sigma}(V_{j,p\alpha} d_{j\sigma}^\dag c_{p\sigma\alpha}+\mathrm{H.c.})
    \end{aligned},
\end{equation}
where now we do not need to specify the Hermitian matrix $h_{ij}$, the bath energies $\ce_{p\alpha}$ and the tunneling matrices $V_{j,p\alpha}$. The goal of the next paragraphs is to identify the dark states of the isolated dot, and this task does not rely on any specific form of the various parameters $h_{ij},\,\ce_{p\alpha},\,V_{j,p\alpha}$. These dark states would then form the slow subspace (i.e. protected from dissipation in a suitable $\gamma_j\to+\infty$ limit) in which the effective Hamiltonian dynamics induced by a nonzero tunneling $V_{j,p\alpha}\neq0$ would take place. We will show that for a suitable choice of the $h_{ij}$ this effective Hamiltonian dynamics is that of higher-spin, possibly many-flavor (depending on the range of $\alpha$) Kondo model.
\par We remind the reader that dark states are eigenstates $\ket{D}$ of the Hamiltonian $H_\textup{dot}$, $H_\textup{dot}\ket{D}=\ce_D\ket{D}$, which are simultaneously annihilated by all jump operators $L_j$, $L_j\ket{D}=0$ \cite{BreuerPetruccione,Rey_Hot_reactive_fermions}. Then, the states $\rho_D=\ketbra{D}$ are stationary states of the dynamics. In our case, following \cite{Rey_Hot_reactive_fermions}, the general construction of the dark states rests only on the strong rotational invariance of the Lindbladian dynamics of the isolated dot:
\begin{equation}\label{eq: dot me}
    \dv{}{t}\rho(t)=-\ii\comm{H_\textup{dot}}{\rho(t)}+\sum_{j\in\textup{dot}}\gamma_j\Big(L_j\rho(t)L_j^\dag-\frac{1}{2}\{L_j^\dag L_j,\rho(t)\}\Big)~,
\end{equation}
in the sense that all components of the total dot spin $S^a=\tfrac{1}{2}\sum_{j,\sigma\tau}(\sigma^a)_{\sigma\tau}d_{j\sigma}^\dag d_{j\tau}$ (where $\sigma^a$ are the Pauli matrices) commute both with the dot Hamiltonian and with all jump operators. Let us diagonalize the Hamiltonian $H_\textup{dot}=\sum_{a\sigma}\ce_a d_{a\sigma}^\dag d_{a\sigma}$ in terms of the appropriate single-particle modes $d_{a\sigma}=\sum_{j}\varphi_a(j)d_{j\sigma}$, where $\varphi_a(j)$ are the eigenfunctions of $h_{ij}$. Then, we can form an $n$-particle polarized state $\ket{\uparrow\uparrow\dots\uparrow}$ in which $n$ fermions with spin $\uparrow$ are put into $n$ different single-body eigenstates of $H_\textup{dot}$ ($n$ must be smaller or equal than the number of dot sites $\ell_d$). This state corresponds to a spin eigenstate $\ket{S,M=S}$ where $S=n/2$, and it is trivially a dark state since all fermions have the same spin projection and thus there cannot be any double occupancies. It will have energy $\ce_{\bm{a}}=\sum_k\ce_{a_k}$, where $\bm{a}=(a_1,\dots,a_n)$ is the set of occupied single-body levels. The rest of the states of the spin-$S$ multiplet are generated by repeated application of the ladder operator for the total spin, $S^-\equiv\sum_j d_{j\downarrow}^\dag d_{j\uparrow}=\sum_a d_{a\downarrow}^\dag d_{a\uparrow}$ Since $\comm{L_j}{S^-}=0$, the states generated in this way are all annihilated by $L_j$ \footnote{Of course, the same set of dark states would be found by starting with the $\downarrow$ polarized state $\ket{\downarrow\downarrow\dots\downarrow}$ in the same levels $\bm{a}$, by repeated application of the raising operator $S^+=(S^-)^\dag$.}. These multiplets of  many-body eigenstates of a Hamiltonian are known as Dicke states \cite{Dicke,PRA_atomic_coh_states,Rey_Hot_reactive_fermions}. We can see that for any non-extremal choice of the number of particles $0<n<\ell_d$ (namely, total spin $S=n/2$) there will be $\binom{\ell_d}{n}$ different multiplets with the same spin quantum numbers but different energies (barring degeneracies of $\ce_{\bm{a}}$ for different $\bm{a}$), where $\binom{\cdot}{\cdot}$ is the binomial coefficient.
\par In general, there is no guarantee that the set of dark states just described exhausts the possible stationary states. In the case of a dot consisting of a linear chain of nearest-neighbor hopping fermions (in either periodic or open boundary conditions), with dissipation acting on every site or on just one, we have verified via exact diagonalization (done with the QuTiP package \cite{qutip1,qutip2}) that this is indeed the case, at least for small chains. For one or two dissipative sites, we have backed up this conclusion with analytical calculations.
\par We notice that the same construction above can be applied to find the stationary state of the full model, i.e. Eq. (1) in the main text. Namely, one needs to diagonalize the quadratic Hamiltonian (a resonant level model) $H=H_d+H_\textup{leads}+H_\textup{tun}=\sum_{a\sigma}\ce_a f_{a\sigma}^\dag f_{a\sigma}$, and then one can construct the set of Dicke states by filling the single-particle $f_{a\sigma}$ modes. Since, as shown before, the resulting states are free of double occupancies, they are dark states for the two-body losses on the dot site, and the true stationary states for finite-sized leads. However, we expect that the dynamics will take a rather long time to bring the state to the Dicke manifold. In \rrefs~\cite{PRL_dissipativeHubbard1D,PRA_liouvilleGapDissipativeHubbard}, it was shown that the minimal decay time (i.e. the inverse of the Liouvillian gap) for a fermionic chain of length $L$ with two-body losses on every site scales as $L^2$. If we restrict the dissipation to act only on one site, the probability of two fermions residing on the dissipative site will be further suppressed, and we can estimate that the longest decay time will scale at least as $L^3$. Since in the NCA that we employed here we take the limit of continuous leads---with infinite particles in them---the Dicke states will never emerge.
\par The eigenstates of the isolated dot dynamics \eqref{eq: dot me} outside the dark subspace---the bright states---will have a finite dissipation rate, and their population will be depleted in time. For now, we assume that there is a parameter regime in which the decay rate of the all bright states can be made sufficiently large. Later, we will show that for more than one dot site, this requirement does not simply boil down to $\gamma\to+\infty$. In the presence of a large decay rate of the bright states, and in the spirit of adiabatic elimination \cite{Garcia-Ripoll_2009} or the dissipative Schrieffer-Wolff transformation \cite{Kessler}, to a first approximation the dynamics will be restricted to the dark subspace, and within this subspace it will be unitary. If we introduce back the coupling to the leads, the latter will mediate transitions between multiplets with neighboring values of $S$. Indeed, we can write
\begin{equation}
    d_{a\sigma}=\sum_{S=0}^{\ell_d/2}\sum_{M=-S}^{S}\sum_{\bm{b}\in \mathcal{A}_S}\delta^{a\sigma}_{SM\bm{b}}\ketbra{S-1,M-\sigma,\bm{b}-(a)}{S,M,\bm{b}}~,
\end{equation}
for certain coefficients $\delta^{a\sigma}_{SM\bm{b}}$. The set $\mathcal{A}_S=\{(a_1,\dots,a_{2S})\vert a_k\neq a_l\}$ is the set of all unordered strings of $2S$ distinct single-particle eigenstates of $H_\textup{dot}$, and the notation $\bm{b}-(a)$ means that the state $a$ is removed from the string $\bm{b}$ (if present). Substituting the above expression in $H_\textup{tun}$ we obtain the Hamiltonian in the dark subspace
\begin{equation}
    \begin{aligned}
        H_\textup{eff}=&\sum_{S=0}^{\ell_d/2}\sum_{M=-S}^{S}\sum_{\bm{a}\in \mathcal{A}_S}\ce_{\bm{a}}\ketbra{S,M,\bm{a}}+\sum_{p\sigma\alpha} \ce_{p\alpha} c_{p\sigma\alpha}^\dag c_{p\sigma\alpha}\\
    &+\sum_{p\alpha\sigma a}\sum_{S=0}^{\ell_d/2}\sum_{M=-S}^{S}\sum_{\bm{a}\in \mathcal{A}_S}(V_{SM\bm{b},p\alpha}^{a\sigma}\ketbra{S,M,\bm{b}}{S-1,M-\sigma,\bm{b}-(a)} c_{p\sigma\alpha}+\mathrm{H.c.})
    \end{aligned},
\end{equation}
with $V_{SM\bm{b},p\alpha}^{a\sigma}\equiv\sum_j V_{j,p\alpha}\varphi_a^\ast(j)(\delta^{a\sigma}_{SM\bm{b}})^\ast$. The above Hamiltonian belongs to the family of the ionic models \cite{Hewson}, that have been extensively studied in the context of magnetic impurities in metals. The low-energy description of such models leads to higher-spin Kondo models in suitable regimes. In general, there will be a lowest-energy multiplet with spin $S^\ast$ (usually, the highest-spin $S^\ast=\ell_d/2$), separated from the next multiplet by an energy gap $\Delta\ce$. If this gap is much larger than the level width induced by the leads, $\Gamma_\alpha\sim \mathcal{N}_{\alpha F}V^2$ (where $\mathcal{N}_{\alpha F}$ is the single-particle density of states at the chemical potential of lead $\alpha$, and $V$ is the typical magnitude of the matrix elements $V_{SM\bm{b},p\alpha}^{a\sigma}$), then $H_\textup{eff}$ can be mapped to a spin $S^\ast$ Kondo model by the usual Schrieffer-Wolff transformation \cite{Hewson,Coleman}. On the other hand, there is no constraint on the number of possible leads, thus opening to the possibility of studying over-screened realizations of these exotic Kondo models, that are known to have non-Fermi liquid ground states \cite{GogolinNersesyanTsvelik, Giamarchi,Hewson,PRB_Kiselev_two_level_Kondo,PRB_Kiselev_multistage_Kondo}. 
\par In any realistic realization of our setup the dissipation rate $\gamma$ will be finite. Then, the coupling to the leads will introduce a residual dissipation, since by repeated tunneling events into the dot sites one ends up in the dissipated states. The quantitative description of the effective dissipation can be achieved by adiabatic elimination of the dissipated modes \cite{Garcia-Ripoll_2009} or, equivalently, by a dissipative Schrieffer-Wolff transformation \cite{Kessler}. It is simple to guess that the maximal-spin multiplet $S_m=   \ell_d/2$ will suffer the largest effective dissipation, since any particle entering the dot will create a double occupancy. Thus, we can expect that the dissipation will couple $S_m$ to $S_m-1$, with a decay rate $\order{V^2/\gamma}$, where $V$ is the tunneling rate to the leads. The higher-lying multiplets will need more tunneling events to reach the dissipative states, with a multiplet of spin $S$ acquiring a dissipative rate only at order $S_m-S+1$ in $V^2$.
\par We now consider the concrete scenario of a dot composed of two dissipative sites, for which the dynamics \eqref{eq: dot me} is amenable to an exact solution. We consider the Hamiltonian 
\begin{equation}
    H_\textup{dot}=\ce_d\sum_{j=0,\sigma}^{1}d^\dag_{j\sigma}d_{j\sigma}-J\sum_\sigma(d^\dag_{0\sigma}d_{1\sigma}+\mathrm{H.c.})=\sum_{a,\sigma}\ce_a d_{a\sigma}^\dag d_{a\sigma}~,
\end{equation}
where the eigenstates are the symmetric and antisymmetric modes $d_{a=\pm,\sigma}=(d_{0\sigma}\pm d_{1\sigma})/2^{1/2}$, corresponding to the energies $\ce_\pm =\ce_d \mp J$. With these states, we obtain the dark subspace multiplet structure shown in \fig~(2) in the main text. In detail, we have the $S=0$ empty dot state $\ket{0}$, the two $S=1/2$ doublets $d_{a\sigma}^\dag \ket{0}$ with energies $\ce_a$ and the $S=1$ triplet $\{d_{+\uparrow}^\dag d_{-\uparrow}^\dag \ket{0},\, 2^{-1/2}(d_{+\uparrow}^\dag d_{-\downarrow}^\dag+d_{+\downarrow}^\dag d_{-\uparrow}^\dag)\ket{0},\,d_{+\downarrow}^\dag d_{-\downarrow}^\dag \ket{0}\}$ with energy $\ce_{S=1}=\sum_a \ce_a=2\ce_d$. We are considering the case $\ce_d<0$, $\abs{\ce_d}>J$ which makes the spin triplet the lowest lying multiplet. In this case, following the construction detailed in the previous paragraph, the effective low-energy theory for a weakly coupled bath would yield a $S=1$ Kondo model. For $-J<\ce_d<0$, the doublet built out of the symmetric states would have the minimum energy $\ce_d-J$, and the effective low-energy Hamiltonian would be the usual $S=1/2$ Kondo model. 
\par We now build the full set of bright states of the $\mathcal{L}_\textup{dot}=-\ii\comm{H_\textup{dot}}{\cdot}+\sum_{j\in\textup{dot}}\gamma_j\big(L_j\cdot L_j^\dag-\frac{1}{2}\{L_j^\dag L_j,\cdot\}\big)$ appearing in \eq~\eqref{eq: dot me}. Let us consider the non-Hermitian part of the Lindblad dynamics, $K=H_\textup{dot}-\ii\gamma/2\sum_j L_j^\dag L_j=H_\textup{dot}-\ii/2\sum_j \gamma_j n_{j\uparrow}n_{j\downarrow}$ (with $n_{j\sigma}=d_{j\sigma}^\dag d_{j\sigma}$ the number operator at site $j$). For a uniform $\gamma_j=\gamma$, this Hamiltonian is a non-Hermitian version of the Hubbard model \cite{Coleman}. Since $K$ conserves the number of fermions (and spin), we can consider each particle number $N_f$ (and spin $S$) sector separately. In the one-particle sector, $K$ coincides with $H_\textup{dot}$, since there are no double occupancies. In the two-particle sectors, it is easy to observe that the eigenstates of $\mathcal{L}_\textup{dot}$ can be inferred from the diagonalization of $K$. Indeed, let $K \ket{r_\alpha}=\kappa_\alpha \ket{r_\alpha}$ be the right eigenstates (with $\ket{l_\alpha}$ the left ones, $\bra{l_\alpha}K=\kappa_\alpha \bra{l_\alpha}$, with the normalization $\braket{l_\alpha}{r_\beta}=\delta_{\alpha\beta}$). If $L_j\ket{r_\alpha}=0$ for all $j$, then $\ket{r_\alpha}$ is one of the dark states built above, and the eigenstates of $\mathcal{L}_\textup{dot}$ in the dark subspace are $\ketbra{r_\alpha}{r_\beta}$, with eigenvalue $\lambda=-\ii(\kappa_\alpha-\kappa_\beta)$ (since dark states are eigenstates of the Hermitian Hamiltonian $H_\textup{dot}$, right and left eigenstates coincide and the $\kappa_\alpha$s are real). This construction of the dark states is valid in all particle-number sectors. Let us consider the $S=0$, $N_f=2$ sector with two fermions of opposite spin. Since each jump operator removes precisely two opposite-spin fermions, we must have $L_j\ket{r_\alpha}=\eta_{j\alpha}\ket{0}$, and we can anticipate that
\begin{equation}\label{eq: eigenm ansatz}
    \sigma=p_0\ketbra{0}+p_1 \ketbra{r_\alpha}{r_\beta}
\end{equation}
will be a right eigenstate of the Lindblad superoperator. Since bright states must be traceless [as $0=\Tr(\mathcal{L}_\textup{dot}\sigma)=\lambda \Tr\sigma$], we have $p_0=-p_1 \braket{r_\beta}{r_\alpha}$. Requiring $\mathcal{L}_\textup{dot}\sigma=\lambda\sigma$ we find 
\begin{equation}\label{eq: lambda}
    \lambda=-\ii(\kappa_\alpha-\kappa_\beta^\ast)~,
\end{equation}
provided that $\sum_j\gamma_j\eta_{j\alpha}\eta_{j\beta}^\ast=-\lambda \braket{r_\beta}{r_\alpha}$. The latter is indeed satisfied: by taking the overlap of $K \ket{r_\alpha}=\kappa_\alpha \ket{r_\alpha}$ with $\ket{r_\beta}$ we find 
\begin{equation}
    \mel{r_\beta}{H_\textup{dot}}{r_\alpha}=\kappa_\alpha\braket{r_\beta}{r_\alpha}+\frac{\ii}{2}\sum_j\gamma_j \eta_{j\alpha}^\ast \eta_{j\beta}~.
\end{equation}
The Hermiticity of $H_\textup{dot}$, $\mel{r_\alpha}{H_\textup{dot}}{r_\beta}^\ast=\mel{r_\beta}{H_\textup{dot}}{r_\alpha}$ yields the desired relation. From the above discussion, we infer that by diagonalizing $K$ in the two-fermion sector we can find the decay rates $-\Re\lambda>0$ by looking at the imaginary parts of the $\kappa_\alpha$s.
\begin{figure}
    \centering
    \includegraphics[width=\linewidth]{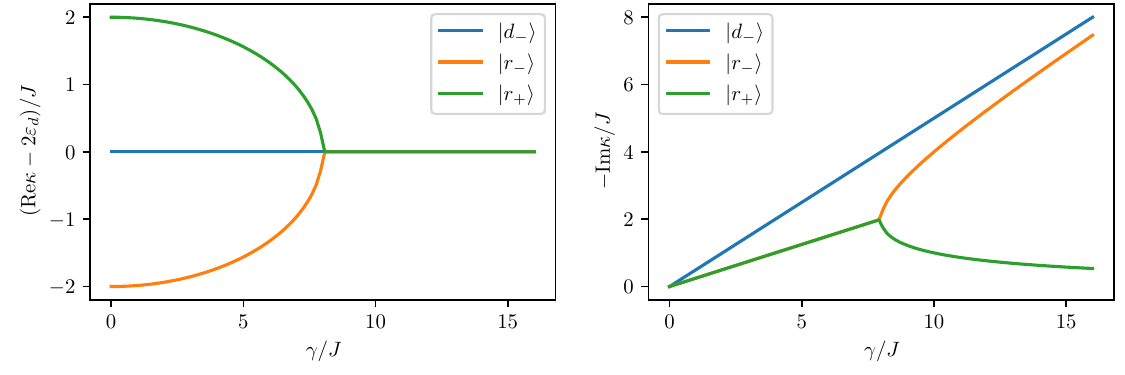}
    \caption{Real (left panel) and imaginary parts (right panel) of the eigenvalues of the non-Hermitian Hamiltonian $K=H_\textup{dot}-\ii\gamma/2\sum_j n_{j\uparrow}n_{j\downarrow}$ in the $N_f=2$, $S=0$ sector, showing the presence of a state ($\ket{r_+}$) which becomes effectively dark for infinite dissipation.}
    \label{fig: bright states}
\end{figure}
\par We now proceed to diagonalize $K$ in the $N_f=2,\,S=0$ sector. We work in real space and we  introduce the basis states \footnote{The “missing” basis state in the subspace with $N_f=2$ and total magnetization $M=0$ is the triplet state $\ket{t}=\frac{1}{\sqrt{2}}(\ket{\uparrow,\downarrow}+\ket{\downarrow,\uparrow})$, which is a dark state belonging to the $S=1$ manifold.} 
\begin{equation}
    \begin{cases}
        &\ket{s}=\frac{1}{\sqrt{2}}(\ket{\uparrow,\downarrow}-\ket{\downarrow,\uparrow})~,\\
        &\ket{d_0}=\ket{\uparrow\downarrow,0}~,\\
        &\ket{d_1}=\ket{0,\uparrow\downarrow}~,
    \end{cases}
\end{equation}
where the notation $\ket{\uparrow,\downarrow}=d_{0\uparrow}^\dag d_{1\downarrow}^\dag\ket{0}$ denotes a state with the $\uparrow$ fermion sits on the $j=0$ site while the other sits at the $j=1$ site, and $\ket{\uparrow\downarrow,0}$ denotes that both fermions are at site $0$ \footnote{We order the labels as $0\uparrow<0\downarrow<1\uparrow<1\downarrow<\dots$.}. In the basis $\{\ket{s},\ket{d_0},\ket{d_1}\}$ the non-Hermitian Hamiltonian $K$ reads
\begin{equation}
    \begin{pmatrix}
    2\ce_d  &   -\sqrt{2}J & -\sqrt{2}J \\
   -\sqrt{2}J &   2\ce_d-\ii\gamma_0/2    &   0\\
    -\sqrt{2}J &   0   &   2\ce_d-\ii\gamma_1/2
    \end{pmatrix}
\end{equation}
Let us consider the simpler case of a uniform dissipation $\gamma_j=\gamma$. Then, we can form the states $\ket{d_\pm}\equiv (\ket{d_0}\pm\ket{d_1})/2^{1/2}$ and we find that $K\ket{d_-}=(2\ce_d-\ii\gamma/2) \ket{d_-}$. Hence, according to \eq~\eqref{eq: lambda}, eigenmatrices of $\mathcal{L}_\textup{dot}$ involving $\ket{d_-}$ will have a decay rate of at least $\gamma/2$. In the remaining subspace spanned by $\{\ket{s},\,\ket{d_+}\}$ we have
\begin{equation}\label{eq: K 2 p s}
    K=
    \begin{pmatrix}
        2\ce_d  &   -2J \\
        -2J     &   2\ce_d-\ii\gamma/2
    \end{pmatrix}~,
\end{equation}
which is easily diagonalized to find the eigenvalues
\begin{equation}
    \kappa_\pm=2\ce_d -\frac{\ii}{4}\gamma\pm\frac{1}{4}(64J^2-\gamma^2)^{1/2}
\end{equation}
and the eigenvectors
\begin{equation}
    \begin{aligned}
        \ket{r_\pm}=[4J^2+(\kappa_\pm-2\ce_d)^2]^{-1/2}
        \begin{pmatrix}
            2J \\
            -\kappa_\pm+2\ce_d
        \end{pmatrix}\\
        \bra{l_\pm}=[4J^2+(\kappa_\pm-2\ce_d)^2]^{-1/2}
        \begin{pmatrix}
            2J,  &   -\kappa_\pm+2\ce_d
        \end{pmatrix}
    \end{aligned}~.
\end{equation}
Since $K=K^T$ is symmetric, $\ket{l_\pm}=(\ket{r_\pm})^\ast$. We sum up the spectral properties of the $N_f=2,\,S=0$ sector in \fig~\ref{fig: bright states}. For the $\ket{r_\pm}$ states we observe two distinct regimes, separated by $\gamma^\ast=8J$. This value of $\gamma$ marks an exceptional point \cite{ReviewNonHermitianUeda} where the two eigenvectors become parallel (and $\kappa_+=\kappa_-$) and $K$ is no longer diagonalizable. For $\gamma<\gamma^\ast$, the two eigenvalues $\kappa_\pm$ have distinct real parts while the imaginary part of both is equal to $-\ii\gamma/4$. In this regime, a larger dissipation rate corresponds to a larger decay rate for the bright states. For larger $\gamma>\gamma^\ast$ the two real parts coincide, while the imaginary parts start to diverge from each other: while $-\Im\kappa_-\sim\ii\gamma/2+\order{J^2/\gamma}$ increases further, $-\Im\kappa_-$ \emph{decreases}, ultimately as $8J^2/\gamma$ for large $\gamma\gg J$. This phenomenon can be seen as an incarnation of the Zeno effect, and implies that for $\gamma\to+\infty$ the $\ket{r_+}$ state (which in this limit coincides with the singlet $\ket{s}$) becomes effectively dark. 
A numerical diagonalization of the Lindbladian with QuTiP confirms that $-2\Im\kappa_+$ is indeed the lowest decay rate at large dissipation. Thus, in order to have an effective dynamics that involves the dark subspace only, we need to increase the intra-dot tunneling $J$ at the same rate as the dissipation $\gamma$, i.e. $J\propto\gamma$. At the same time, to have an appreciable Kondo temperature for the effective $S=1$ (or $S=1/2$) Kondo model at low energy we must keep the gap $\Delta\ce=\abs{\ce_d+J}$ between the two lowest-lying multiplets finite: this requirement implies that $\ce_d$ must be scaled proportionally to $\gamma$ as well. In other words, the adiabatic elimination of the bright states is possible only in the regime $\abs{\ce_d}\sim J\sim \gamma\gg\Gamma$, where $\Gamma$ is the level width induced by the leads. See also the companion paper \cite{companion} for further discussions of these points.
\par  To complete our description of the bright states, let us consider the completely filled state $\ket{F}=\ket{\uparrow\downarrow,\uparrow\downarrow}$, which is an eigenstate of $K$: $K\ket{F}=(4\ce_d-\ii\gamma)\ket{F}$. The corresponding eigenmatrix of $\mathcal{L}_\textup{dot}$ can be constructed by analogy with \eq~\eqref{eq: eigenm ansatz}, namely as a superposition of $\ketbra{F}$, $\ketbra*{r_\alpha^{(2,0,0)}}{r_\beta^{(2,0,0)}}$ and $\ketbra{0}$, where $\ket*{r_\alpha^{(N_f,S,M)}}$ indicates a right eigenstate of $K$ in the sector with $N_f$ particles, spin $S$ and magnetization $M$ \footnote{Since $\ket{F}$ has $S=M=0$ and the jump operators conserve spin, the whole eigenmatrix must be composed only of states with the same spin quantum numbers.}. We will not reproduce the full calculation here, but it is easy to understand that the real part of the eigenvalue of $\mathcal{L}_\textup{dot}$ will receive a contribution of $-\Im(\kappa_\alpha+\kappa_\beta)$ from each term $\ketbra{r_\alpha}{r_\beta}$. Hence, the resulting decay rate will be at least $2\gamma$---indeed, the it is the state with the faster decay rate.
\par Finally, we consider the $N_f=3$, $S=1/2$ sector. Since all states in this sector have one unpaired spin, and by particle-hole correspondence with the $N_f=1$, $S=1/2$ states, we expect to find two $S=1/2$ doublets. Indeed, we can think of these states as featuring a “hole” on top of the full state $\ket{F}$: we have verified that the states $\ket{N_f=3,a\sigma}\equiv d_{-a\sigma}\ket{F}$ are the sought eigenstates, with eigenvalues $\kappa=4\ce_d-\ce_{-a}+\ii\gamma/2$.
\par We remark that in the present case the diagonalization of $K$ in all symmetry sectors allows to reconstruct all eigenmatrices of the Lindblad superoperator by suitably combining the projectors $\ketbra*{r_\alpha^{(N_f,S,M)}}{r_\beta^{(N_f^\prime,S^\prime,M^\prime)}}$, $\ketbra*{r_\alpha^{(N_f-2,S,M)}}{r_\beta^{(N_f^\prime-2,S^\prime,M^\prime)}}$ and $\ketbra{0}$. As observed above, each term will add a positive contribution $-\Im\kappa$ to the decay rate $-\Re\lambda$. Since we have proven that the $3$- and $4$-particle sectors have all $-\Im\kappa\propto\gamma$, any time an eigenmatrix features an eigenstate of $K$ with more than $2$ particles appears, the decay rate will be linear in $\gamma$. The upshot of this discussion is that the lowest decay rate, vanishing for $\gamma\to+\infty$, only comes from the $N_f=2$, $S=0$ sector considered before.
\begin{figure}
    \centering
    \includegraphics[width=\linewidth]{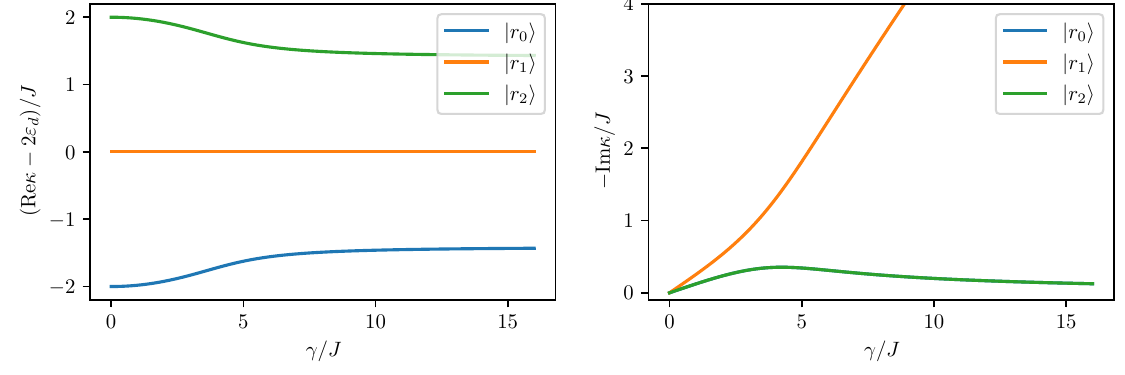}
    \caption{Real (left panel) and imaginary parts (right panel) of the eigenvalues of the non-Hermitian Hamiltonian $K=H_\textup{dot}-\ii\gamma/2 n_{0\uparrow}n_{0\downarrow}$ for the case of inhomogeneous dissipation in the $N_f=2$, $S=0$ sector, showing the presence of two states ($\ket{r_{1,2}}$) which become effectively dark for infinite dissipation.}
    \label{fig: bright states single diss}
\end{figure}
\par The conclusions drawn for the homogeneously dissipated case $\gamma_j=\gamma$ are essentially unchanged also in the completely inhomogeneous one with all dissipation concentrated only on one site, $\gamma_0=\gamma$, $\gamma_1=0$. The dark subspace does not change, while the non-Hermitian Hamiltonian in the $N_f=2$, $S=0$ subspace becomes:
\begin{equation}
K=
    \begin{pmatrix}
    2\ce_d  &   -\sqrt{2}J & -\sqrt{2}J \\
   -\sqrt{2}J &   2\ce_d-\ii\gamma/2    &   0\\
    -\sqrt{2}J &   0   &   2\ce_d
    \end{pmatrix},
\end{equation}
whose eigenvalues have to be found numerically. The results are shown in \fig~\ref{fig: bright states single diss}. As in the homogeneous case, there is one mode with $\Re\kappa_1=2\ce_d$ and a monotonic decay rate $-\Im\kappa\sim\gamma/2$, but both the other modes have the same non-monotonic imaginary part---which is generally smaller than in the homogeneous case. The decay rates of these states rises linearly $-\Im\kappa\sim \gamma/8$ for small dissipation, reaches a maximum for $\gamma\approx4J$ and then slowly decreases as $-\Im\kappa\sim 2J^2/\gamma$ for large dissipation. In this setup, there is no exceptional point. An exceptional point appears instead in the $N_f=3$ particle sector, in which 
\begin{equation}
    K=
    \begin{pmatrix}
        3\ce_d  &   J \\
        J     &   3\ce_d-\ii\gamma/2
    \end{pmatrix}
\end{equation}
in the basis $\{d_{0\sigma}\ket{F},\,d_{1\sigma}\ket{F}\}$ has the same form of \eq~\eqref{eq: K 2 p s} \footnote{The matrix reported is identical for the two possible  magnetizations $M=\pm1$.}. We see that for a single-site dissipation, also the $N_f=3$ sector has a state with a non-monotonic decay rate, reaching a maximum for $\gamma=4J$. The decay rate in this sector is always larger than the one in the $N_f=2$, $S=0$ sector, albeit they tend to coincide for large dissipation. The full state $\ket{F}$ has a monotonic decay rate, $K\ket{F}=(4\ce_d-\ii\gamma/2)\ket{F}$, which is obviously smaller than in the homogeneous case. This brief analysis of the inhomogeneous setup allows to draw two conclusions: First, a single dissipative site is sufficient to provide all states outside the dark subspace with a finite dissipation rate, and so to implement the mapping to higher-spin Kondo models described above. The second conclusion is that this setup is more fragile than the homogeneous one, in the sense that more bright states become essentially dark for large dissipation $\gamma\gg J$, and that the overall decay rates are smaller than in the homogeneous case.
\par The ideas presented above can be extended to larger dot sizes, and we expect that the minimal decay rates of the bright states will have a non-monotonic behavior. We have explicitly verified this expectation for $\ell_d=3$. Thus, in general, the fine-tuning $J\sim\abs{\ce_d}\sim\gamma$ will be needed. This fine-tuning might be circumvented by choosing a different geometry for the dot sites, as for instance a potential well hosting more than one bound state.
\bibliography{biblioDissKondo.bib}